\documentclass[manuscript]{aastex}

\received{9/12/2014}
\accepted{4/4/2015}

\shorttitle{MIR diagnostic diagram for BLAGNs} 

\shortauthors{Coziol et al.}

\begin{document}

\title{Comparing narrow and broad-line AGNs, in a new diagnostic diagram for emission-line galaxies based on WISE data}
\author{R. Coziol\altaffilmark{1},  J. P. Torres-Papaqui\altaffilmark{1}, H. Andernach\altaffilmark{1} }

\affil{Departamento de Astronom\'{\i}a, Universidad de Guanajuato, Apartado Postal 144, 36000, Guanajuato, Guanajuato, M\'exico}

\begin{abstract} 

Using a new color-color diagnostic diagram in the mid infrared built from WISE data, the MIRDD, we compare narrow emission-line galaxies (NELGs) that 
exhibit different activity types (star-forming galaxies, SFGs, and AGNs, i.e.,LINERs, Sy2s and TOs), as determined using one standard diagnostic diagram in the 
optical (BPT-VO), with broad-line AGNs (QSOs and Sy1s) and BL Lac objects at low redshift ($z \le 0.25$). We show that the BL Lac objects occupy in the 
MIRDD the same region as the LINERs, whereas the QSOs and Sy1s occupy an intermediate region, between the LINERs and the Sy2s.In the MIRDD these 
galaxies trace a sequence that can be reproduced by a power law, $F_\nu = \nu^{\alpha}$, where the spectral index, $\alpha$, varies from 0 to $-2$, 
which is similar to what is observed in the optical-ultraviolet part of the spectra of AGNs with different luminosities. 

For the NELGs with different activity types, we perform a stellar population synthesis analysis, confirming that their specific positions in the MIRD depend on their star formation histories (SFH), and demonstrating that the ${\rm W}2-{\rm W}3$ color is tightly correlated with the level of star formation in their host galaxies. In good agreement with the SFH analysis, a comparison of their MIR colors with the colors yielded by energy distributions (SEDs) of galaxies with different activity types, shows that the SED of the LINERs is similar to the SEDs of the QSOs and Sy1s, consistent with AGN galaxies with mild star formation, whereas the SEDs of the Sy2s and TOs are consistent with AGN galaxies with strong star formation components. For the BL Lac objects, we show that their blue MIR colors can only be fitted with a SED that has no star formation component, consistent with AGNs in elliptical-type galaxies. 

From their similarities in MIR colors and SEDs, we infer that, in the nearby universe, the level of star formation activity most probably increases in the host galaxies of emission-line galaxies with different activity types along the sequence BL~Lac$\rightarrow$LINER$\rightarrow$QSO/Sy1$\rightarrow$Sy2$\rightarrow$TO$\rightarrow$SFG.

\end{abstract}


\section{Introduction}
\label{INTRO}

A large fraction of galaxies in the nearby universe show emission lines in their nucleus
\citep{Heckman80,Kennicutt92,Ho96,HFS97,York00,Stoughton02,TP11}. However, only a few percent of these galaxies exhibit
broad lines similar to quasars, where we are sure that the nuclear activity is due to the accretion of gas onto a
supermassive black hole (SMBH) at the center of their host galaxies \citep{Lynden-Bell69,Soltan82,RawlingsSaunders91}.

In the 1980s, different studies demonstrated that by comparing the ratios of different spectral emission lines in the
optical it is possible to devise standard diagnostic diagrams that allow to identify the dominant source of ionization
of the gas in narrow emission-line galaxies (NELGs). In their seminal paper, \citet[][BPT]{BPT81} only recognised three
possible sources: massive young stars in Star Forming Galaxies (SFGs), the accretion of gas onto a SMBH in Active
Galactic Nuclei (AGNs), and shock-wave heating due to supernovae winds or their remnants. Later on, \citet[][VO]{VO87}
argued based on empirical evidence that only the two first mechanisms proposed by BPT are most likely to explain the line ratios in NELGs, and
proposed simple criteria to distinguish between SFGs and AGNs in standard diagnostic diagrams. In particular, one
diagnostic diagram that compares the line ratio [NII]$\lambda6583$/H$\alpha$ with the line ratio
[OIII]$\lambda5007$/H$\beta$---hereafter, the BPT-VO diagram---proved to be one of the most useful tool to make this
distinction \citep[e.g.][]{Stasinska06}.

\begin{figure} 
\epsscale{0.9} 
\plotone{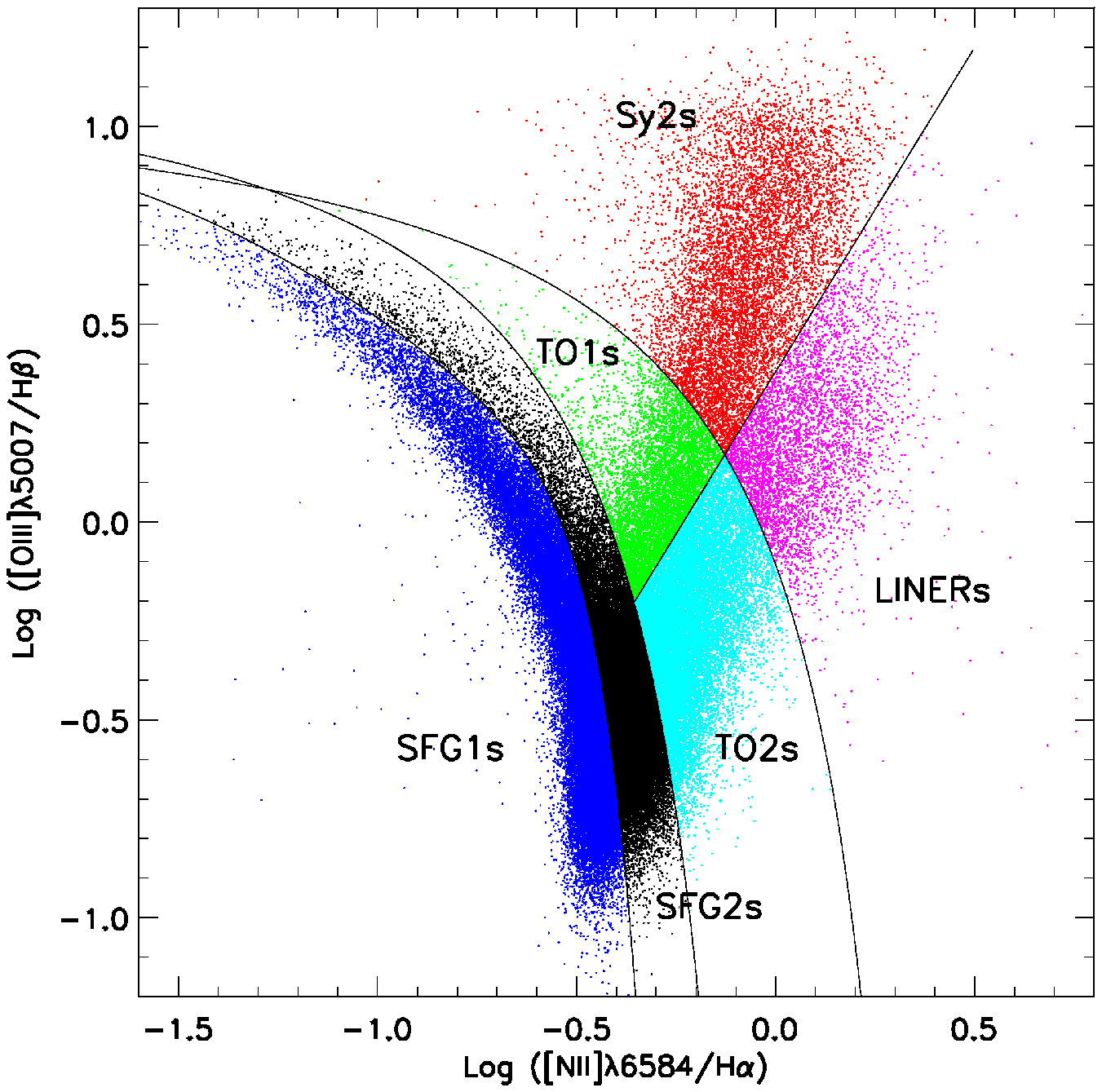} 
\caption{BPT-VO diagram for the NELGs selected for our MIR study. The
boundaries between the NELGs with different activity types are those defined in \citet{TP13}. 
\label{BPTVO}}
\end{figure}

Thanks to large spectroscopic surveys like the Sloan Digital Sky Survey \citep[SDSS;][]{York00,Abazajian09}, we have now
the possibility to apply the BPT-VO diagram to very large samples of NELGs at low redshift. What is observed, then, is a
continuous sequence that takes the shape of the greek symbol $\nu$\ \citep{Kew01,Kau03,TP12a}, the SFGs forming the left
branch and the AGNs the right one (cf., Fig.~\ref{BPTVO}).

The nature of the sequence produced by the SFGs in the BPT-VO diagram is now well understood \citep[see][and references
therein]{Coz96,Coz99}. This is due to a variation in spiral galaxies of the metallicity of the gas
\citep{McCall85,DE86,TP11,TP12b}: the ratio [NII]$\lambda6583$/H$\alpha$ rises with the metallicity, while the ratio
[OIII]$\lambda5007$/H$\beta$ decreases as the level of excitation of the gas diminishes with the temperature. However,
the explanation for the AGN branch is more ambiguous and controversial \citep{Viegas92,Cooke00,Nagao06}. The most
natural assumption is that the high [NII]$\lambda6583$/H$\alpha$ ratio seen in AGNs---their distinctive trait in the
BPT-VO diagram---reflects a high abundance of chemical elements. But this would imply that the gas metallicity in AGNs is much higher than in SFGs, ignoring the cooling effect of metals in interstellar gas: as the
metallicity grows, the gas temperature decreases due to cooling by emission, reducing the ratio
[OIII]$\lambda5007$/H$\beta$, and not increasing it as it is observed on the right branch of the $\nu$-shape distribution
formed by AGNs \citep[see the discussion in][]{Coz11}. Alternatively, many authors have suggested that AGNs display an
excess of nitrogen for their metallicity \citep{Osterbrock70,Storchi89,Storchi91,Hamann93,Coz99,Coz11,TP11,TP12a}. But
this would imply that the formation of SMBHs at the center of galaxies and the formation of stars in their bulge are coupled processes, the two structures developing at the same time
\citep{Dokuchaev91,HamannFerland92,BoyleTerlevich98,Franceschini99,Haiman04,Menanteau05,LouJiang08,Haiman07,Li07,Bennert08,Fanidakis11,Treister11,TP12a,TP12b,Floyd13,TP13}.

Another difficulty in explaining the AGN branch in the BPT-VO diagram is related to the Low ionization Nuclear
Emission-line Regions (LINERs) and the Transition-type Objects (TOs), which constitute the majority of narrow-line AGNs
in the nearby universe \citep{Heckman80,HFS93,Ho96,Kew01,HFS03,Kau03,Kau09}. That LINERs are genuine AGNs is suggested
by the extreme position of these galaxies in the BPT-VO diagram: together with Seyfert~2 galaxies (Sy2s), LINERs form a
continuous sequence at the right end of the $\nu$-shape branch (c.f., Fig.~\ref{BPTVO}), implying, according to
photoionization models, that the ionizing source in a LINER is the same as in a Sy2, but intrinsically weaker in
luminosity \citep{OD83,Willner85,GK86,HFS93,Schmitt97,Barth98,Larkin98,Ho99,Sarzi05,Starling05,Kew06,Maoz07,Masegosa11}. However, and despite the BPT-VO diagram, many researchers still doubt the AGN nature of LINERs, suggesting, instead,
that they are either ionized by shocks \citep{Heckman86,Forbes93,Veilleux95,Kim98,Veilleux99}, or by very massive stars
\citep{Shields92,Dopita95,Maoz98}. Some authors have even suggested that the gas in LINERs is ionized by stars much less
massive than O and B stars, but much hotter: the so-called post-asymptotic giant branch stars (PAGBs). This appellation,
unfortunately, is being used vaguely in the literature, as it could refer to pre-planetary nebulae (the genuine PAGBs),
planetary nebulae (PNs), or even hot white dwarfs
\citep{DC71,Taniguchi00,Terashima00,Flohic06,S08,CF11,Giraud11,Singh13,Papaderos13}.

Determining the nature of TOs based on the BPT-VO diagram is also problematic, because star formation is very intense in
these galaxies, while it is much lower in the LINERs \citep{TP13}. This has led \citet{Kew01} to propose new boundaries in the BPT-VO diagram between AGNs and SFGs, suggesting that TOs form a buffer zone (cf.,
Fig.~\ref{BPTVO}), where star formation is still the predominant source of ionization of the gas. Consequently, the most
accepted interpretation for TOs seems to be that they are AGNs hidden behind extended regions of intense star formation
happening near their center \citep{HFS93,HFS97,VGV97,GVV99,HFS03,TP12a,TP13}. 

On the other hand, explaining why
the TOs differ from the LINERs and the Sy2s is not obvious, because all these galaxies trace an apparently continuous
sequence in the BPT-VO diagram. One possible explanation is that TOs are AGNs at an earlier stage of evolution, whence the name
``transition-type objects'' , since they would be expected to eventually transform into either Sy2s or LINERs when the level of star formation
would have decreased sufficiently \citep{OS88,Wu98,Maoz99,Vila-Vilaro00,Sarzi05,Lee07,Haan08,WW08,GD08,Constantin08,Constantin09,Chen09,Netzer09,Montero-Dorta09,Chen10,Carpineti12,TP13}.

The ambiguities as to the nature of LINERs and TOs, as inferred from the BPT-VO diagram, is a serious drawback in our
understanding of the AGN phenomenon. It suggests that standard diagnostic diagrams based on emission-line ratios could
be misleading, and that conclusions on the nature of AGNs that were classified using this method are possibly erroneous.
Moreover, because we cannot apply the BPT-VO diagram to BLAGNs, establishing a connection between the different AGN
types observed in the nearby universe is a complicated issue. What is needed, therefore, are alternative means by which
we can verify independently the classification of the activity of NELGs as obtained using the BPT-VO diagram, and that
could be used to compare NELGs with BLAGNs directly, i.e., by comparing the same characteristics. This can now be done
thanks to the Wide-field Infrared Survey Explorer \citep[WISE;][]{Wright10}.

WISE is an all-sky photometric survey in four wave bands in the mid infrared (MIR): 3.4 $\micron$ (W1), 4.6 $\micron$ (W2), 12 $\micron$ (W3) and 22 $\micron$
(W4). The WISE All-Sky Release Source Catalog, which we use in this study, contains astrometry and photometry
for 563,921,584 objects detected on the Atlas Intensity Images \citep{Cutri13}. This survey has already produced 
important results, showing, in particular,  that emission-line galaxies with different activity types have distinctive
MIR characteristics \citep[e.g.][]{Jarrett11,Stern12,Mateos12,Donoso12,Clemens13,Assef13,Yan13}. Following the same line
of research, \citet{Coz14} have recently devised, using WISE data, a new diagnostic diagram, the MIRDD, that allows to
separate empirically NELGs with different activity types, in a way that is fully consistent with the classification
obtained from the BPT-VO diagram. The MIRDD consists in comparing the ${\rm W}2-{\rm W}3$ colors with the ${\rm
W}3-{\rm W}4$ colors, and although the separation criteria are empirical, it is suspected, as it is also the case for
the BPT-BVO diagram, that the reason why such segregation is possible in the MIR is because these two colors vary
systematically as the contributions of the two main components forming the spectral energy distribution (SED) in the
MIR, an AGN and star formation in the host galaxies, change with the activity type.

For example, in the MIRDD the AGN nature of LINERs seems now much more compelling. This is because their MIR colors are
significantly different from those of SFGs and PAGBs, implying that although massive stars or white dwarfs might still
be assumed to ionize the gas in LINERs, these stars obviously cannot explain their MIR emission, whereas an AGN can explain both.
Moreover, the LINERs and Sy2s trace in the MIRDD a continuous sequence that is consistent with the color sequence
produced by power laws with different spectral indexes, in good agreement with photoionization models
\citep[e.g.,][]{OF06,Maoz08}, whereas SN shock-wave is not a mechanism that is capable of heating dust. Also, the ${\rm
W}2-{\rm W}3$ colors of the NELGs in the MIRD were found to become gradually redder along the sequence
LINER$\rightarrow$Sy2$\rightarrow$TO$\rightarrow$SFG, which suggests that this color is mostly sensitive to the present
level of star formation activity in the host galaxies of emission-line galaxies \citep{Donoso12,TP12a,TP13}.

Consequently, we here propose a new study that should allow us to better understand how the MIRDD works, and gain new
information on how BLAGNs are connected with AGNs in NELGs. This article is organized in the following way. We introduce our
sample of nearby ($z \le 0.25$) NELGs and BLAGNs in Section~\ref{DATA}, and use
the BPT-VO diagram to separate the NELGs into six different activity types. In Section~\ref{MIRDD_NELG}, we discuss how these galaxies are differentiated in the MIRDD, and support this discussion by doing a statistical analysis of the MIR colors. In Section~\ref{BEff} we examine what are the effects on our statistical analysis of
having different distributions in redshift and different number of resolved galaxies in our sample.
In Section~\ref{SFvsAT}, we perform a new stellar population
synthesis analysis on the NELGs \citep[e.g][]{Coz11,TP12a,Plauchu12,TP13}, deducing their star formation histories
(SFHs),  to check how well this characteristic can explain the positions of galaxies with different activity types
in the MIRDD.  We then proceed by studying the relation in NELGs between the ${\rm W}2-{\rm W}3$
color and the intensity of star formation in their host galaxies. In Section~\ref{SED_NELGs}, we complement the SFH
analysis by studying how the changes of colors in the MIRDD are related to the different components that are forming the SEDs of
emission-line galaxies with different activity types. Finally, in Section~\ref{MIRDD_BLAGN}, we compare the MIRDD of
BLAGNs and BL~Lac objects with the MIRDD of NELGs, and in Section~\ref{SED_BLAGNs} we compare their SEDs. In our
discussion, Section~\ref{DISCUSSION}, we consider the consequences of our comparison of BLAGNs with NELGs in the MIR,
and summarize our results and conclusions in Section~\ref{CONCLUSION}.

\section{Samples and MIR data} 
\label{DATA}

The NELGs used to compare the BPT-VO diagram with the MIRDD were selected from the main spectroscopic catalog of the
Sloan Digital Sky Survey Data Release~7 \citep[SDSS DR7;][]{Abazajian09}, which represents the completion of the SDSS-II
project. The SDSS DR7 catalog includes the spectra of $9.3 \times 10^5$\ galaxies, $1.2 \times 10^5$ quasars and $4.6
\times 10^5$\ stars\footnote{http://www.sdss.org/dr7/start/aboutdr7.html}. Only the objects identified as galaxies were
considered for our study, keeping all those that have a redshift $z \le~0.25$. After correcting for the redshifts,
stellar population templates produced by STARLIGHT \citep{Cid05} were subtracted from the data, leaving pure emission
spectra. For our study, we selected only the NELGs that have optical emission line ratios, [NII]$\lambda6583$/H$\alpha$
and [OIII]$\lambda5007$/H$\beta$, with a signal-to-noise ratio ${\rm S/N} \ge 3$, and ${\rm S/N} \ge 10$ in the adjacent
continuum \citep[][]{Brinchmann04}, which amounts to 229,618 NELGs~\citep{TP12a}.

To retrieve the MIR data, we cross-correlated the positions of the NELGs, as found in SDSS DR7, with the positions of
the objects in the WISE All-Sky Release Source Catalog. This was done using the X-Match pipeline in VizieR
\citep{OBM00}, applying a search radius of five arcseconds around the position of each galaxy
\citep[e.g.,][]{Clemens13}. For each match, the algorithm determines the angular distance from the candidate to the target. 
By plotting this angular distance, we found distributions that peak around 0.2 arcsecond, with relatively very few galaxies ($\sim 1$1\%) with a distance larger than
1~arsecond. By reviewing the WISE identification, we eliminated all the duplicate matches (which represent much less that $1$\% of the candidates), keeping only those that have the smallest angular differences. Then, selecting only the NELGs that have WISE fluxes with ${\rm S/N} > 3$ (corresponding to
quality flags, {\it ph\_qual}, equal to A or B) in all of the four WISE MIR bands, and that have a contamination and
confusion flag that is clear ($cc\_flags = 0$), we obtained 133,095 NELGs (i.e., 58\% of our original spectroscopic
sample).

Using the BPT-VO diagram, we classified our sample of MIR NELGs according to their activity type. The results are shown
in Fig.~\ref{BPTVO}. The different activity types are those defined in \citet{TP13}. The NELGs are separated in three
main types, the SFGs (66.7\% of the MIR NELGs), the TOs (22.7\%), the Sy2s (7.0\%) and the LINERs (3.6\% ). Our MIR sample of nearby
NELGs clearly favors galaxies where star formation is relatively high \citep[similar to what was found
by][]{Donoso12}. We also separated the SFGs and TOs in two subtypes. Compared to the SFG1s (56\% of the SFGs), the SFG2s
are star forming galaxies that show an excess of nitrogen due to starburst winds \citep[see explanations in][]{TP12b},
whereas the TO2s (74\% of the TOs) differ from the TO1s because of their lower level of excitation, which makes them
more similar to LINERs than Sy2s \citep{TP13}. Our final WISE selected, NELGs sample is thus composed of 49,724 SFG1s,
39,026 SFG2s, 7,752 TO1s, 22,671 TO2s, 9,361Sy2s and 4,741 LINERs.

The BLAGNs in our sample come from the 13th edition of the catalog of quasars and active nuclei published by
\citet{VCV10}, which is also accessible through VizieR. From this catalog, we selected all the entries identified as
QSOs or Sy1s (more specifically for the Sy1s, only those identified as S1 in the catalog). Although it has become
customary to separate quasars into type~1 and type~2, we have made no such distinction in our study, because this would have implied
doing a followup in X-rays \citep[e.g.][]{Derry03}, which is beyond the scope of the present study. Also, although BL Lac
objects are not BLAGNs (since their spectra show no emission lines), we have included them in our study because, due to
their non-thermal continua in the optical, they are presumed to represent the ``purest'' AGN cases possible, and,
consequently, they can be important points of reference in the MIRDD \citep[e.g.,][]{Plotkin12,Massaro13}.

Like for the NELGs, we restricted the redshift of our candidates to $z \le~0.25$, obtaining three initial
spectroscopic samples of 7,038 Sy1s, 3,510 QSOs, and 169 BL Lac objects. The MIR data from the WISE All-Sky Release Source
Catalog were retrieved in the same way as for the NELGs. Comparing the angular distances we found wider distributions than for the NELGs, with a maximum near 1~arcsecond (compared to 0.2 arcsecond for the NELGs), but with only a few percent of
candidates with angular distances larger than 2~arcseconds. After eliminating all the duplicated matches, and applying the same selection
criteria as the NELGs for the MIR fluxes, we ended up with 5,077 Sy1s, 2,323 QSOs, and 55 BL Lac, which represent,
respectively, 72\%, 66\% and 33\% of the initial spectroscopic samples.

\begin{figure} 
\epsscale{0.9} 
\plotone{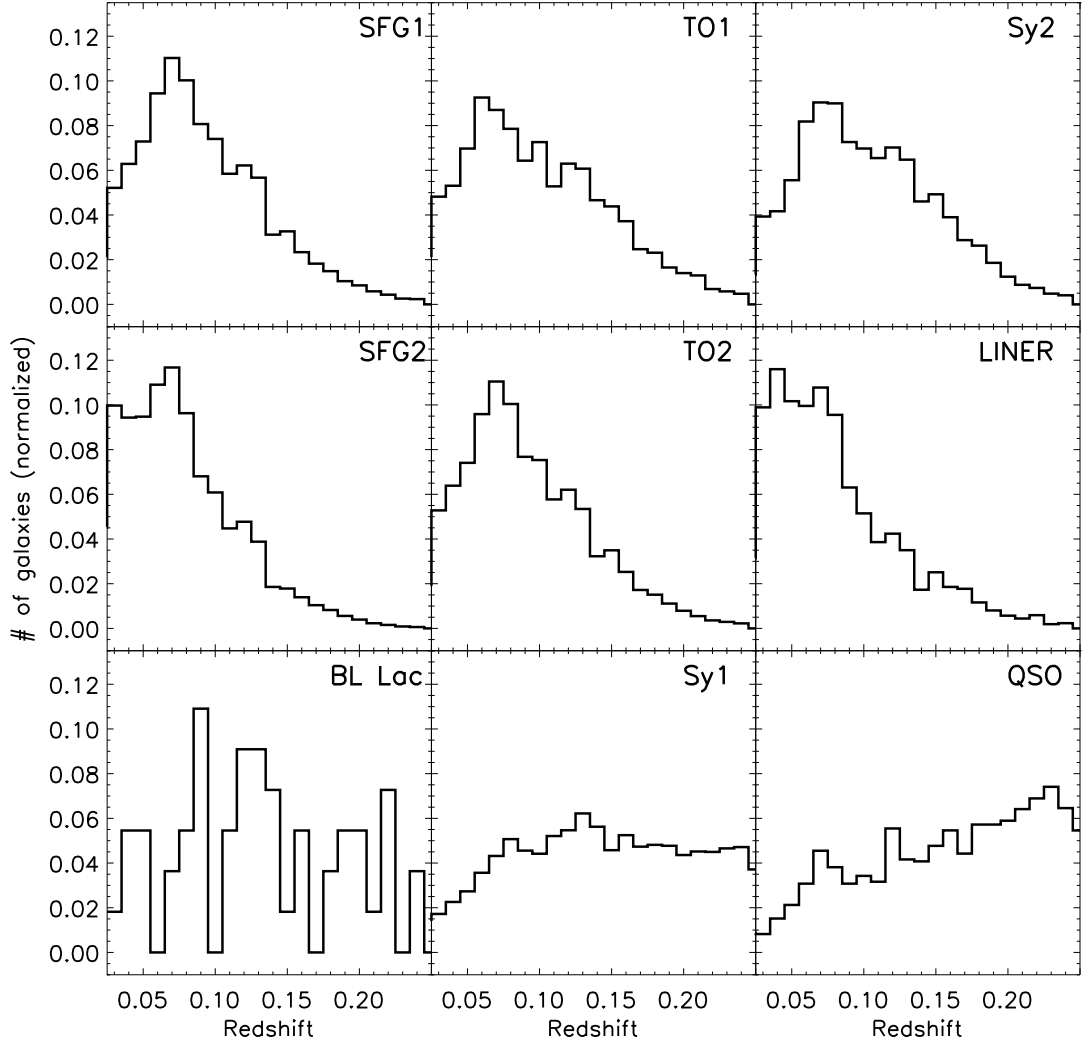} 
\caption{Histogram showing the normalized  redshift distributions of the different emission-line galaxies in our sample. 
\label{Hz}}
\end{figure}

In Fig.~\ref{Hz} we show the normalized distributions in redshift of all the selected emission-line galaxies, as
separated by activity types. As it should have been expected, the NELGs are better represented at low redshifts than the
BLAGNs. The distributions of the SFGs and TOs are similar, showing a maximum at $z\simeq 0.07$ and a long exponentially
decreasing tail, that stays well populated up to $z\simeq0.14$ for the two SFGs and TO2s, but decreasing less rapidly with the redshift in the case of
the TO1s. The SFG2s also differ from the SFG1s, presenting a higher fraction of galaxies at low redshifts. The
distribution for the Sy2s is similar to the distributions of the TO1s. The
distribution for the LINERs is similar to the distribution of the SFG2s, showing an almost constant fraction of
galaxies from $z=0$ to $z=0.07$, the fraction of galaxies decreasing rapidly at higher redshift. For the BQSOs we see the
inverse trend, their fraction increasing with the redshift. The QSO distribution reaches its maximum only at $z=0.25$,
while the distribution of the Sy1s reaches its maximum at much lower redshift, their fractions remaining constant over the range
$z=0.08$ to $z=0.25$. The difference between the Sy1s and QSOs is as expected from the classification criterion, which is  
based on a difference in absolute magnitude: consequently, the fraction of BLAGNs classified as QSOs instead of Sy1s is naturally expected to increase with the redshift. The BL Lac objects show no obvious trend with the redshift. But,  this is most probably due to small statistics (there is only 55 objects in this sample). 

From the different redshift distributions it is expected that a greater number of NELGs will be spatially resolved in WISE.
Indeed, revising the values of the flag {\it ext\_flags}, we find that 66\% of the LINERs are resolved, compared
to 28\%, 35\% and 23\% of the Sy2s, TOs and SFGs, respectively. Similarly, we find that a high fraction of BL Lac
objects are also resolved, $\sim 47\%$, which is relatively high considering their small number and homogeneous
distribution in redshift. In comparison, only 17\% of the Sy1s and 9\% of the QSOs are resolved. In the next section, we
will examine in more detail the possible consequences of these differences on our analysis.

\section{Results}

\subsection{Tracing the MIRDD for nearby NELGs with different activity types} 
\label{MIRDD_NELG}

\begin{figure} 
\epsscale{0.5} 
\plotone{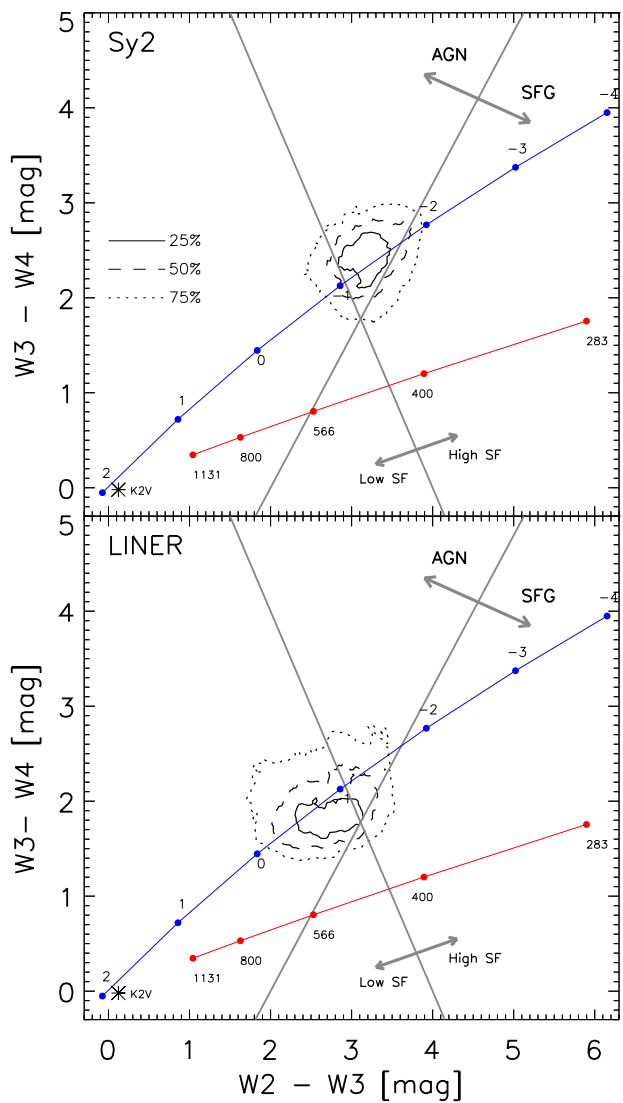} 
\caption{MIRDD for the Sy2s and LINERs. The distributions 
correspond to normalized density contours around the most probable colors. The line with a positive slope separates AGNs
from SFGs and the one with a negative slope separates NELGs according to the level of star formation in their host galaxies. The upper blue curve is the color sequence traced by a power law with different spectra indexes, and the lower
red curve is the color sequence for a black body at various temperatures \citep[Table 1 in][]{Wright10}. The color for a typical K giant is also
indicated as an asterisk at the bottom left \citep[also in][]{Wright10}. 
\label{mirdd_Sy2LINER}} 
\end{figure}

\begin{figure} 
\epsscale{0.5} 
\plotone{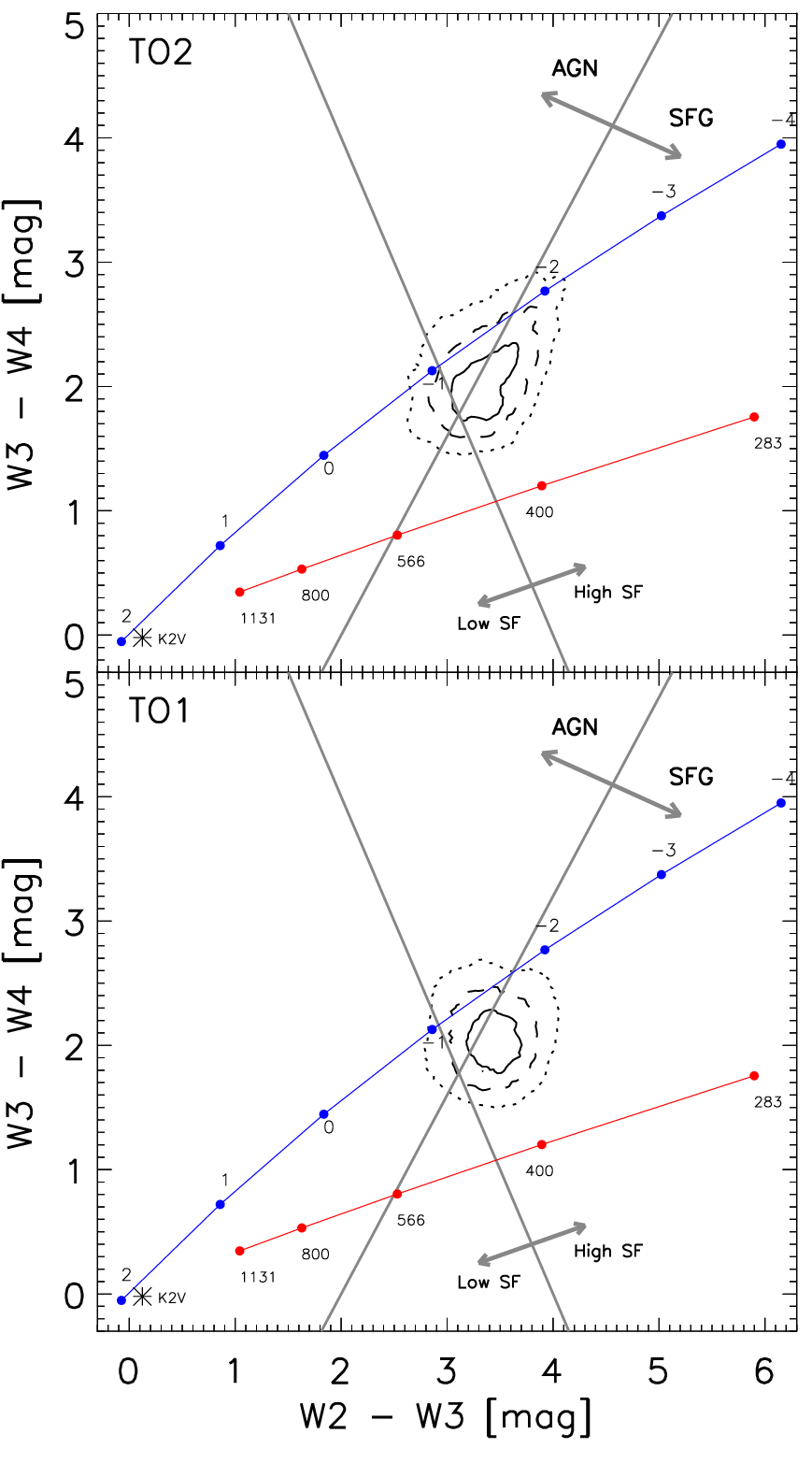} 
\caption{MIRDD for the TOs. The boundaries and curves are as described in Fig.~\ref{mirdd_Sy2LINER}. 
\label{mirdd_TO}} \end{figure}

\begin{figure} 
\epsscale{0.5} 
\plotone{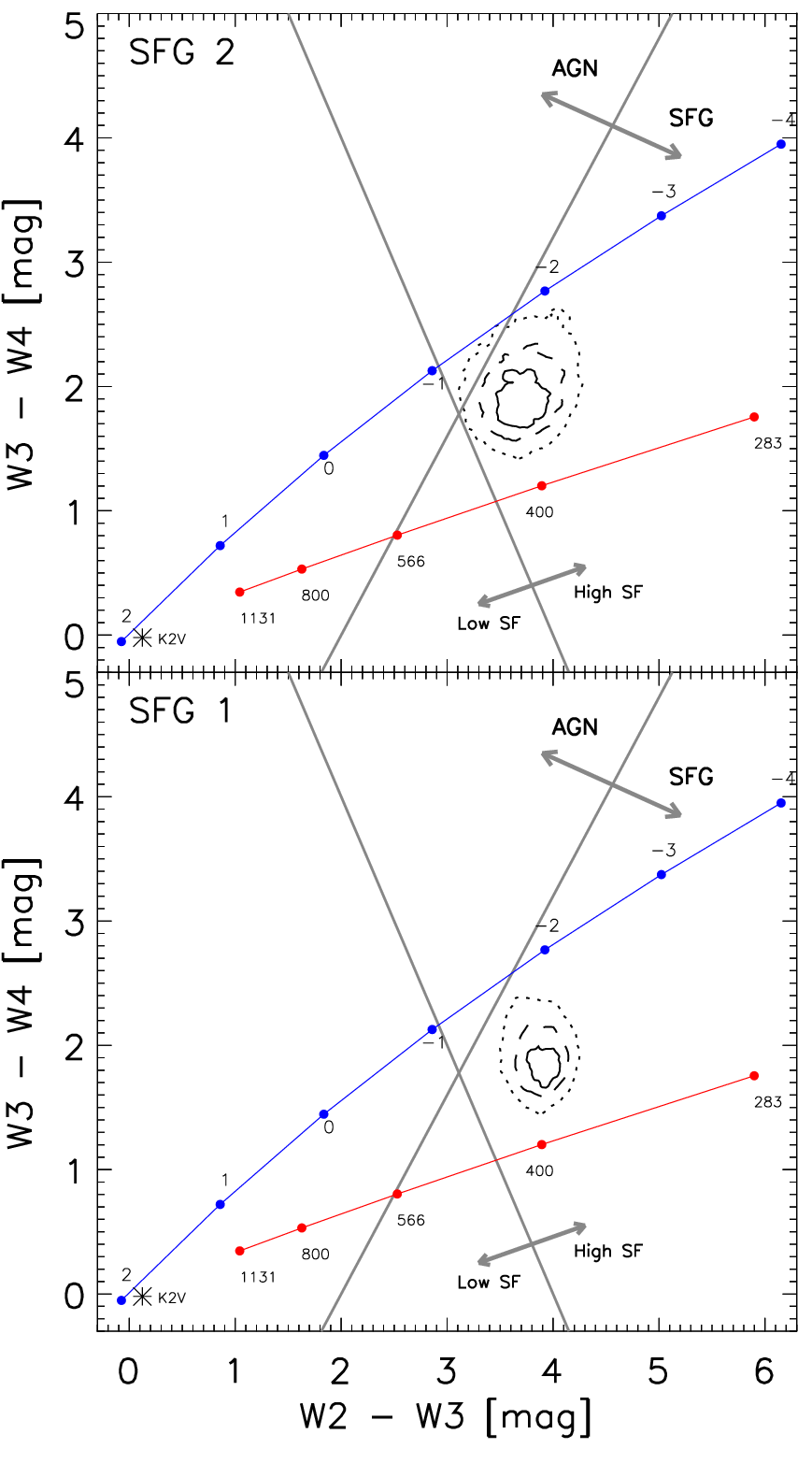} 
\caption{MIRDD for the SFGs. The boundaries and curves are as described in Fig.~\ref{mirdd_Sy2LINER}.
\label{mirdd_SFG}} 
\end{figure}

We show the MIRDD for the NELGs in Fig.~\ref{mirdd_Sy2LINER} for the Sy2s and the LINERs, in Fig~\ref{mirdd_TO} for the TOS and in Fig.~\ref{mirdd_SFG} for the SFGs. The different regions of the MIRDD were
determined empirically in \citet{Coz14}. The analytical relations identifying these different regions are the following:
if $x =$ ${\rm W}2-{\rm W}3$, and $y=$ ${\rm W}3-{\rm W}4$, then the separation between AGNs and SFGs is given by the
expression $y = 1.6x -3.2$, and the separation between high and low star formation activity is given by the expression
$y = -2.0x +8.0$. One can see that on the AGN side we find mostly Sy2s and LINERs, whereas on the side of low star formation activity the majority of the NELGs are LINERs.  

The MIRDD and the BPT-VO diagrams yield highly consistent results for the LINERs and Sy2s. In both diagrams these two
AGNs trace a continuous sequence, the Sy2s differing from the LINERs in the optical by their higher level of ionization,
and, consequently in the MIR, by their slightly higher dust temperatures. Moreover, we find that the colors of the
LINERs and Sy2s are consistent with the colors produced by different power laws, $F_\nu = \nu^{\alpha}$, the
spectral index, $\alpha$, varying between 0 and $-1.5$ in the LINERs and between $-1$ and $-2$ in the
Sy2s. As they are, these different values are consistent with what is observed in the optical-ultraviolet region of the spectra of AGNs
with different luminosity \citep[e.g.][]{OF06,Maoz08}.

Note that the position of the LINERs in the MIRDD is different from the position of the LINERs as found in Figure~12 in
\citet{Wright10}, which has been taken as reference by many authors in the literature. The reason for such a difference
is because the LINERs in the \citet{Wright10} sample are ultra-luminous infrared galaxies (ULIRGs), which are quite
different from the ``standard'' LINERs as defined in the optical by \citet{Heckman80}. This difference is well
documented in \citet{Coz96}, where it is shown that what distinguishes these two types of LINERs is their different
levels of star formation. Indeed, as one can verify, the ULIRGs classified as LINERs in  \citet{Wright10} occupy in the MIRDD the same position as the TOs and SFGs in Fig.~\ref{mirdd_SFG}. It is important to stress this difference, because it demonstrates that although
emission-line galaxies may have similar spectral characteristics in the optical, their MIR colors, and their ${\rm W}2-{\rm W}3$
colors in particular, may differ significantly, depending on the different levels of star formation in their host galaxies. 

Similarly, the different positions occupied by the Sy2s and LINERs in the MIRDD are consistent with a higher level of
star formation in Sy2s than in LINERs \citep[][]{Lawrence85,Dultzin88,TP13,Rosario13,Coz14}. This may also explain
their different power laws, the steeper spectral index for the Sy2s implying an excess in IR emission, which could be due to star formation  \citep{TP12b}. 

The MIRDDs for the TOs in Fig~\ref{mirdd_TO} is also in good agreement with the BPT-VO diagram. In both diagrams,  
the TOs seem to occupy a sort of buffer zone between the SFGs and AGNs \citep{Kew01}. The fact that this is due to star formation is more obvious in the MIRDD, where the ${\rm W}2-{\rm W}3$ colors for the TOs are significantly distinct from those of the LINERs (contrary to the continuous sequence shown in the BPT-VO diagram).  The MIRDD of the TOs clearly emphasizes the high levels of star formation in their host galaxies \citep{Kew01,TP12a,TP13}. 

The differences between the TOs and SFGs are better appreciated by comparing Fig.~\ref{mirdd_TO} with
Fig.~\ref{mirdd_SFG}. The SFG1s have slightly redder ${\rm W}2-{\rm
W}3$ colors and bluer ${\rm W}3-{\rm W}4$ colors than the SFG2s. The highest similarities are
between the TO1s and SFG2s, which, again, is in good agreement with the relative positions occupied by these galaxies in
the BPT-VO diagram. As expected, the colors of the SFGs are clearly distinct from those produced by a power law. On the other hand, none  fall on the black body curve. 

\begin{figure} 
\epsscale{0.7} 
\plotone{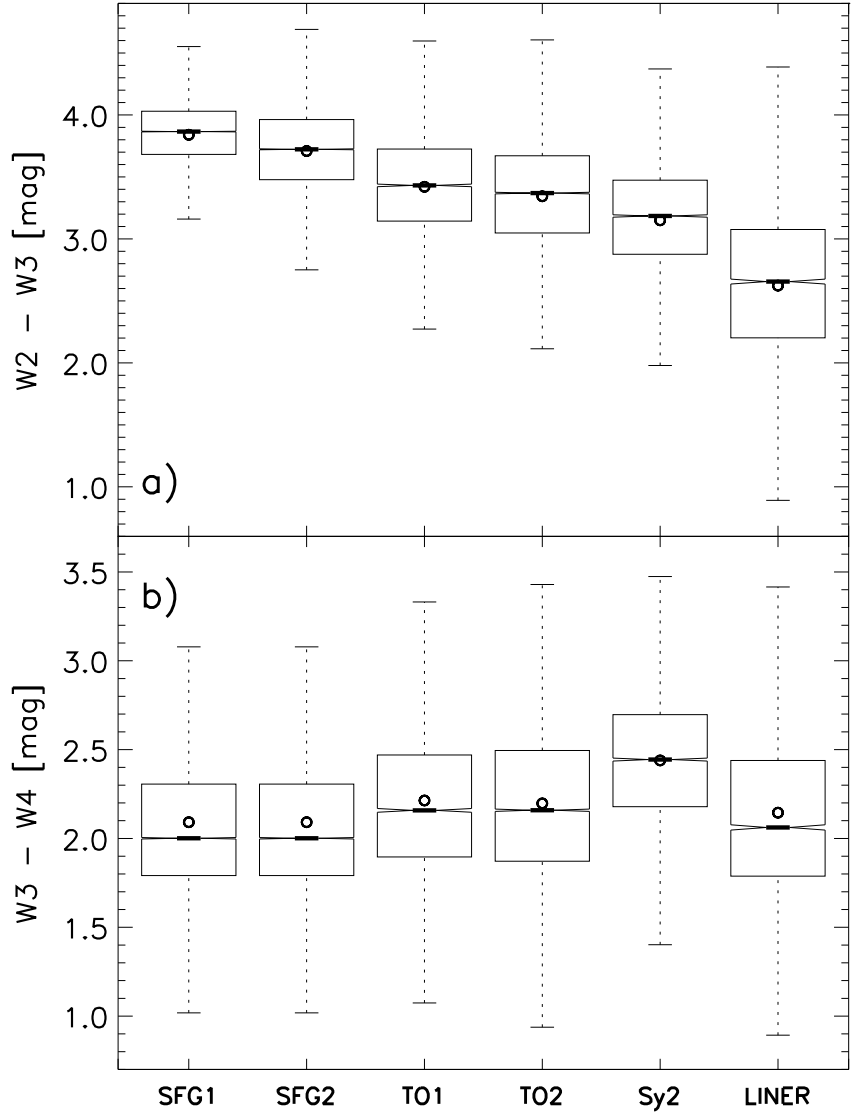} 
\caption{Box-whisker plots of the MIR colors of the NELGs with different
activity types. The lower side of the box is the 25th percentile, $Q_1$, and the upper side is the 75th percentile,
$Q_3$. The whiskers correspond to $Q_1 - 1.5\times IQR$ and $Q_3+ 1.5\times IQR$, where $IQR$ is the interquartile
range. The median is shown as a bar and the mean as a circle. Although barely visible, the box-whisker plots include
notches, which are V-shape regions, drawn around the medians, that have a width proportional to the interquartile range:
$\pm 1.58 \times IQR/\sqrt{N}$, where $N$ is the size of the sample. Samples with overlapping notches have comparable medians. 
\label{bw_stdNELGs}} 
\end{figure}

In general, we distinguish in the MIRDD a definite trend for the ${\rm W}2-{\rm W}3$ color to become redder along the
sequence LINER$\rightarrow$Sy2$\rightarrow$TO2$\rightarrow$TO1$\rightarrow$SFG2$\rightarrow$SFG1. This trend is obvious
in Fig.~\ref{bw_stdNELGs}a, where we present the box-whisker plots for the MIR colors of the different NELGs in our
sample. These plots include notches around the medians that are not overlapping, which suggests that the medians are significantly different. This is not the case for the ${\rm W}3-{\rm W}4$ colors shown in Fig.~\ref{bw_stdNELGs}b. The SFG1s and SFG2s have similar medians, and so have the TO1s and TO2s. There is, consequently, a weaker trend in ${\rm W}3-{\rm W}4$ colors, the NELGs becoming bluer, by only 0.5 mag, along the sequence  Sy2$\rightarrow$TO$\rightarrow$SFG. The LINERs do not follow this sequence, being slightly bluer than the SFGs. 

\begin{figure} 
\epsscale{1} 
\plotone{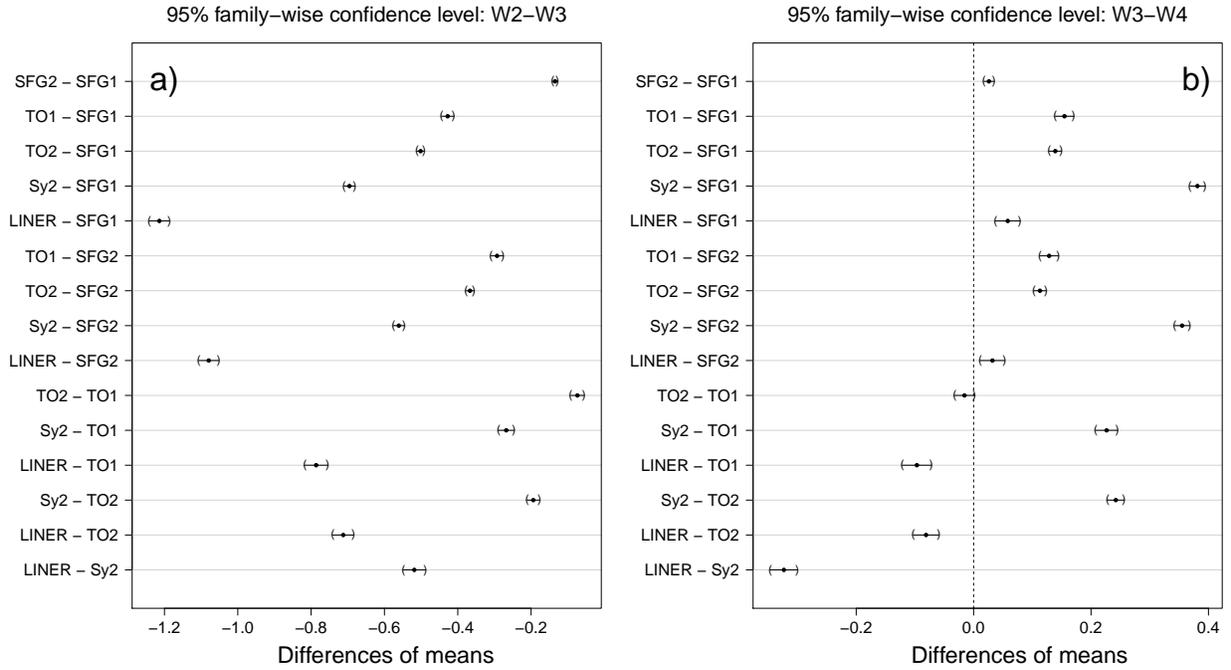} 
\caption{Results of the max-t tests in terms of the 95\% confidence intervals. This parametric test compares simultaneously the differences of mean MIR colors between each pair of subsamples (family-wise). A confidence interval including zero
indicates no statistically significant difference in colors. Negative color differences suggest that the first subsample of galaxies in the
pair compared are on average bluer than the second subsample of galaxies. 
\label{ci_stdNELGs}}
\end{figure}

These differences in colors are confirmed statistically in Fig.~\ref{ci_stdNELGs}, where we present the simultaneous
95\% family-wise confidence intervals, as obtained using the max-t test \citep{HBW08,HSH10}. A more explicit explanation
of this parametric statistical test, and how to interpret its results, can be found in \citet{TP13}. In summary, the
test compares the mean MIR colors, family-wise, i.e., comparing simultaneously all pairs of subsamples possible. For
example, in Fig.~\ref{ci_stdNELGs}a the test for the ${\rm W}2-{\rm W}3$ colors confirms that the LINERs are bluer by $\sim 1.2$ mag, than the SFG1s, while the smallest differences are found between the TO2s and
TO1s (smaller than $0.1$). In Fig.~\ref{ci_stdNELGs}b, the test also confirms that the differences between the mean colors
are much smaller in ${\rm W}3-{\rm W}4$ (fluctuating around zero by $\sim 0.3$ mag) than in ${\rm W}2-{\rm W}3$. 

\subsubsection{Effect of the resolved galaxies on our statistical analysis} 
\label{BEff}

Our statistical analysis clearly shows that the ${\rm W}2-{\rm W}3$ color is a very sensitive parameter in the MIRDD,
varying monotically by $\sim 1.2$ mag along the sequence
LINER$\rightarrow$Sy2$\rightarrow$TO2$\rightarrow$TO1$\rightarrow$SFG2$\rightarrow$SFG1. This suggests that the ${\rm
W}2-{\rm W}3$ color must be related to some intrinsic physical characteristic that is systematically varying with the
activity type of the galaxies. On the other hand, we have also found that our samples contain different fractions of
resolved galaxies at different redshifts. In particular, our sample of LINERs turned out to include a significantly
larger number of resolved galaxies at lower redshifts than the other samples. Thus, it seems important to examine what
is the importance of these differences on the results of our statistical analysis.

According to the WISE explanatory supplement\footnote{http://wise2.ipac.caltech.edu/docs/release/allsky/expsup/}, the
photometry for resolved sources, as identified by the flag {\it ext\_flags}, underestimates their fluxes. Physically,
this means that for the resolved galaxies only their inner regions are included in the photometric aperture. And even
more so, if the galaxies are at low redshift. On the other hand, because the contamination and confusion flags are clear
({\it cc\_flags}$ = 0$), and the quality flags ({\it ph\_qual}) are equal to A or B for all the resolved galaxies, we
are sure that the data in these regions are of good quality. What is left to establish, therefore, is the effect on our
statistical analysis of having only partial fluxes for the resolved galaxies. 

To determine this effect, we have compared the
colors of the unresolved galaxies ($ ext\_flags = 0$) with those of the resolved galaxies ($ext\_flags \ne 0$). What we
found, is that, in general, i.e., independently of the activity type, underestimating the fluxes makes the MIR colors bluer. There
are two possible causes for this ``blueishness effect'' (BEff). One is an observational bias: in the MIR, older stellar
populations have typical blue colors \citep[e.g., the colors of a K-type star in the MIRDD; see
also][]{Wright10,Plotkin12,Jarrett13}, and since stellar populations in the center of galaxies tend to be older, on
average, than in its periphery, then, the resolved galaxies are expected to be bluer than 
the unresolved galaxies. The other cause is physical: because galaxies with different morphologies do not form their stars in
the same way \citep[e.g.,][]{Coz11,TP13}, then galaxies with early-type morphologies will be expected
to have bluer colors, on average, than galaxies with later-type morphologies. 

To distinguish between the two causes described above, we have first quantified the BEff comparing in each activity-type
samples the mean colors of the unresolved galaxies with the means for the total samples. We found that the BEff for the
${\rm W}2-{\rm W}3$ and ${\rm W}3-{\rm W}4$ colors are of the order of 0.35 and 0.32 mag for the LINERs, compared to
0.04 and 0.06 mag for the Sy2s, 0.05 and 0.07 mag for the TO1s, 0.09 and 0.10 mag for the TO2s, 0.02 and 0.00 mag for
the SFG1s, and 0.04 and 0.03 mag for the SFG2s. For comparison sake, the BEff for the BL Lac objects are 0.16 and 0.01,
and practically zero for the Sy1s and QSOs. So, only the LINERs (and BL Lac objects) are significantly affected. This
implies that the BEff cannot be explained by an observational bias, but that LINERs must be physically different from
the other galaxies.

\begin{figure} 
\epsscale{0.8} 
\plotone{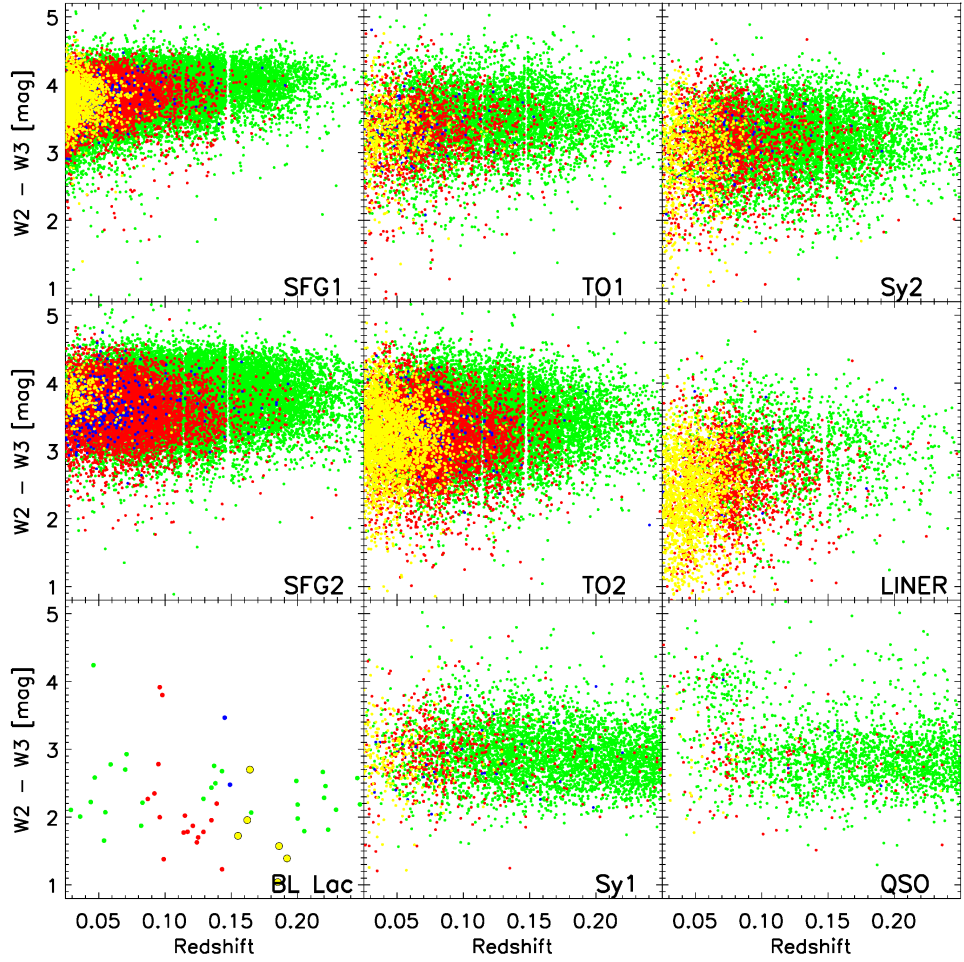} 
\caption{Variations with redshift of the ${\rm W}2-{\rm W}3$ colors of galaxies with different activity-types and levels of resolution, according to the WISE photometry flag ($ext\_flags$): 0 (green), 1 (red) , [2,3,4], (blue), and 5 (yellow).  The gaps at $z \sim 0.115$ and $z \sim 0.148$ for the NELGs are  observational artefacts in SDSS. 
\label{w2w3vszvsext}}
\end{figure}

To examine this point in more detail, we compare in Fig.~\ref{w2w3vszvsext} the distributions of the ${\rm W}2-{\rm
W}3$ colors, which is the parameter that shows the largest variations, as a function of the redshift. In each
activity-type samples the galaxies were separated into four groups (0, 1, [2,3,4] and 5), as given by the WISE photometry
flag ($ext\_flags$): where 0 indicates a point source, and the degree of resolution increases from 1 to 5. As expected,
we see that the fraction of resolved galaxies increases at low redshift. However, the BEff becomes more important at low
redshift only in the LINERs, which confirms that this effect is caused by a physical difference. 

\begin{figure} 
\epsscale{0.8} 
\plotone{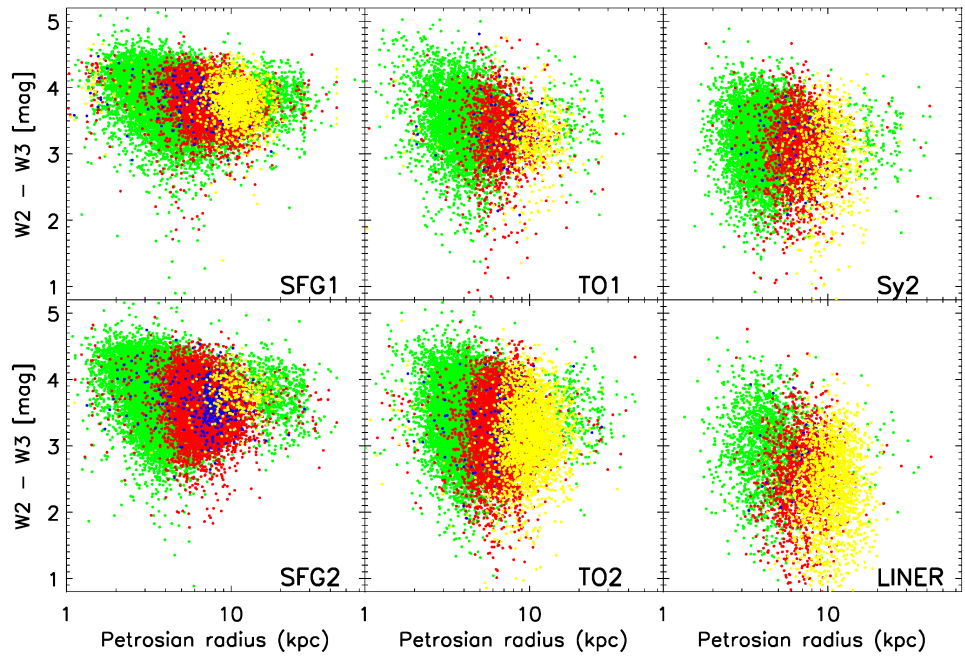} 
\caption{Variations with Petrosian radius of the ${\rm W}2-{\rm W}3$ colors of NELGs with different activity-types and levels of resolution.  The colors have the same meaning as in Fig.~\ref{w2w3vszvsext}.
\label{w2w3vsPRvsext}}
\end{figure}

To support this interpretation, we show in Fig.~\ref{w2w3vsPRvsext} for the NELGs the distributions of their ${\rm
W}2-{\rm W}3$ colors as a function of their Petrosian radii in kpc, as measured in the SDSS $z$\ band \citep[assuming a pure
Hubble flow, with $H_0 = 75$ km s$^{-1}$ Mpc$^{-1}$;][]{TP12a}. It can be seen that the level of resolution increases
with the physical size of the NELGs. In general, the LINERs contain more extended galaxies than the other NELGs, and the
BEff systematically increases with their size. Moreover, for comparable ranges in redshift and comparable ranges in
galaxy size the LINERs always show bluer colors. This can only be explained assuming, on average, older stellar
populations in the LINERs than in the other NELGs \citep{TP12a,TP13}. Consequently, this result supports our statistical
analysis, suggesting that the differences in ${\rm W}2-{\rm W}3$ colors between NELGs with different activity-types are
related to different stellar populations in their host galaxies.

\begin{figure} 
\epsscale{0.8} 
\plotone{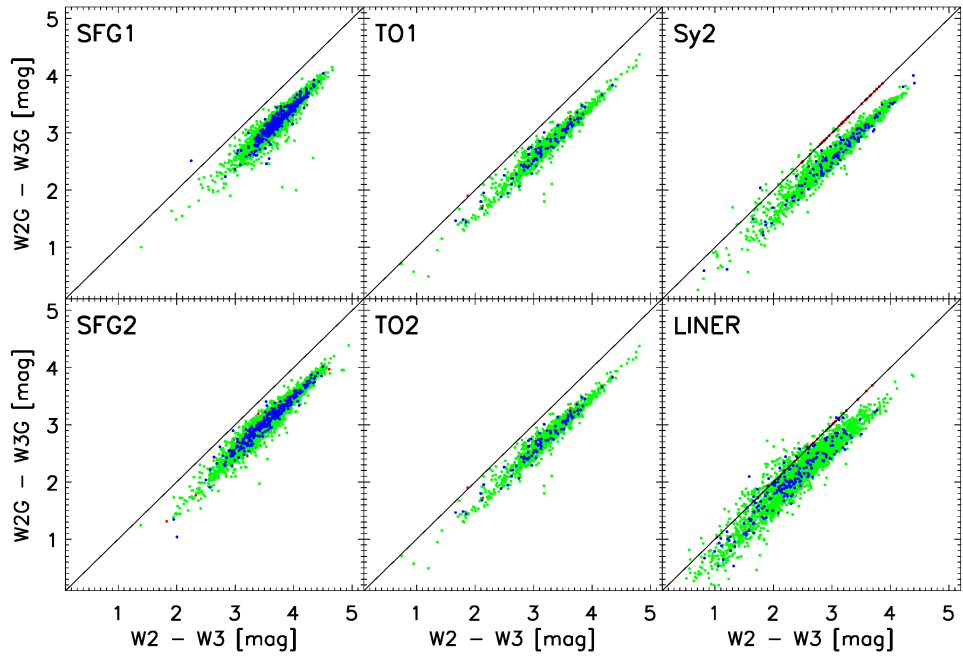} 
\caption{Comparison of  the ${\rm W}3-{\rm W}4$ colors with the ${\rm W}3G-{\rm W}4G$ colors, with different WISE photometry flag $w4gflg$ values: 0 (red), 1 (green) , 17 (blue).
\label{w3w4vsw3gw4g}}
\end{figure}

Also according to the WISE explanatory supplement, it is suggested that in the case of resolved galaxies one should use the
magnitudes obtained by integrating the fluxes using the photometry apertures for 2MASS (i.e., using $w?Gmag$ instead of
$w?mag$). To verify how this would affect our statistical analysis, we have consequently retrieved the available
aperture photometry measurements in the WISE All-Sky survey using the GATOR catalog query service at
IPAC/IRSA\footnote{http://irsa.ipac.caltech.edu/about.html}. We have found data for 83\% of the resolved LINERs, 67\% of
the Sy2s, 57\% of the TO1s, 66\% of the TO2s, 47\% of the SFG1s, sand 55\% of the SFG2s. For comparison sake, we also
retrieved the data for the BL Lac objects and BLAGNs, and found magnitudes for 67\% of the resolved Sy1s, 59\% of the
QSOs and 31\% of the BL Lac. Comparing the colors, we have found that the trend for the resolved galaxies to become bluer
significantly increases. This is shown in Fig.~\ref{w3w4vsw3gw4g}, where we compare the colors obtained using $w?Gmag$
(${\rm W}2G-{\rm W}3G$) with the colors using $w?mag$. One can see also that this effect does not depend on the different photometry flag (we used $w4gflag$).  

Using the $w?Gmag$, the BEff for the ${\rm W}2G-{\rm W}3G$ and
${\rm W}3G-{\rm W}4G$ colors would be 0.52 and 0.41 mag for the LINERs, 0.57 and 0.24 mag for the Sy2s, 0.62 and 0.24 mag for
the TO1s, 0.60 and 0.25 mag for the TO2s, 0.58 and 0.19 mag for the SFG1s, and 0.60 and 0.25 mag for the SFG2s.
Similarly, we found a BEff of 1.1 and 0.24 mag for the BL Lac, 0.43 and 0.25 for the Sy1s, and 0.62 and 0.29 for the
QSOs. It is interesting to note that the resolved BLAGNs show a similar effect as the NELGs, suggesting that even in
luminous AGNs, the MIR colors may be sensitive to a variation in stellar populations. In particular, 
using the $w?Gmag$ we have found a few BL Lac objects to have negative colors, which is typical of inactive, elliptical galaxies, which is as expected from what is known a priori of these galaxies \citep{Kotilainen98,Scarpa00,Falomo14}.

In general, taking into account that the fraction of resolved galaxies varies significantly with the sample, we have
calculated (using a weighted mean) that only the LINERs would be significantly affected, becoming
bluer, in both colors, by $\sim$0.2 mag compared to the other galaxies. Consequently, using the $w?Gmag$ instead of
$w?mag$ would change nothing to our statistical analysis.

\subsection{Variation of star formation in NELGs with different activity types} 
\label{SFvsAT}

In the previous section, we have shown that the systematic variation of ${\rm W}2-{\rm W}3$ color of NELGs with different activity types must be related to a variation in their physical characteristics. In \citet{Coz14} it was suggested that this
characteristic is the present level of star formation in the host galaxies of the NELGs. This was also suggested by
other authors \citep[e.g.,][]{Mateos12,Jarrett13,Rosario13}. In particular, \citet{Donoso12} have
shown that the ${\rm W}2-{\rm W}3$ is a valuable indicator of star formation activity, when no other data are
available. This suggests that the ${\rm W}2-{\rm W}3$ color sequence,
LINER$\rightarrow$Sy2$\rightarrow$TO2$\rightarrow$TO1$\rightarrow$SFG2$\rightarrow$SFG1, traced by the NELGs with
different activity types in the MIRDD, may be due to an increase in the level of star formation in their
host galaxies \citep{TP12a,TP13}.

It is possible to verify directly this interpretation for the ${\rm W}2-{\rm W}3$ color sequence by applying a stellar
population synthesis analysis using STARLIGHT to the SDSS spectra of the NELGs \citep[e.g.,][]{TP13}. The stellar
population synthesis method consists in quantifying in the optical spectra of galaxies the contributions of stellar
populations with different luminosities or masses. One can then transform these contributions in terms of star formation
rates (SFRs), and obtain in this way a snapshot of how a galaxy formed its stars over its lifetime, which constitutes
its star formation history (SFH). Further details on how the SFH is obtained with STARLIGHT can be found in
\citet{Coz11}, \citet{TP12a} and \citet{Plauchu12}. Note that because the SDSS spectra cover mostly the inner parts of the
galaxies, this legitimates a comparison of the SFHs with the colors obtained using only $w?mag$ in WISE \citep{TP12b}.

\begin{figure} 
\epsscale{0.8} 
\plotone{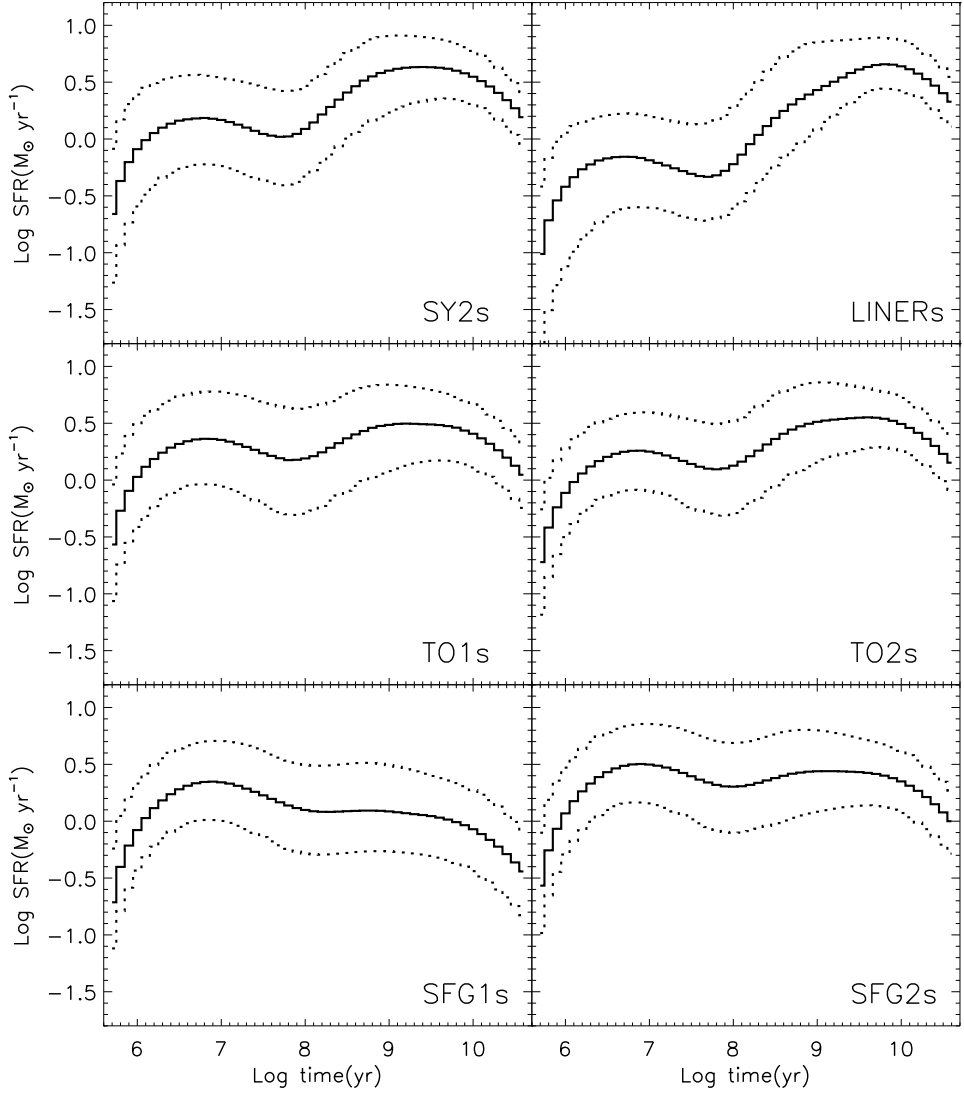} 
\caption{Variations of the SFH in NELGs with different activity types at  different time. The abscissa gives the age of the stellar populations in the library used for our analysis.
The continuous curves are the median SFRs and the two dashed curves are the first and third quartiles.
\label{sfh_NELG}}
\end{figure}

\begin{figure} 
\epsscale{0.7} 
\plotone{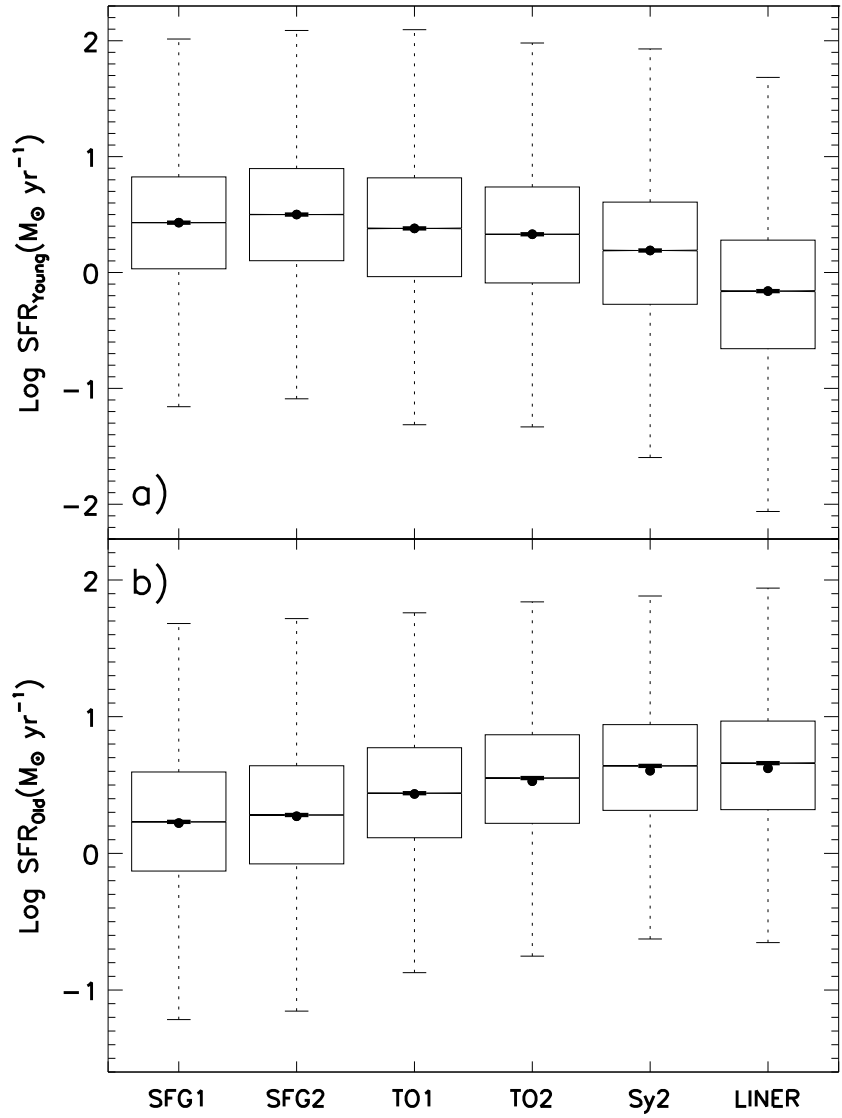} 
\caption{Box-whisker plots showing the variations of the two characteristic SFRs in
NELGs with different activity types: a) SFR$_{\rm Young}$, the most recent maximum in star formation rate, and b)
SFR$_{\rm Old}$, the maximum star formation rate in the distant past. 
\label{SFR_NELG}} 
\end{figure}

\begin{figure} 
\epsscale{1} 
\plotone{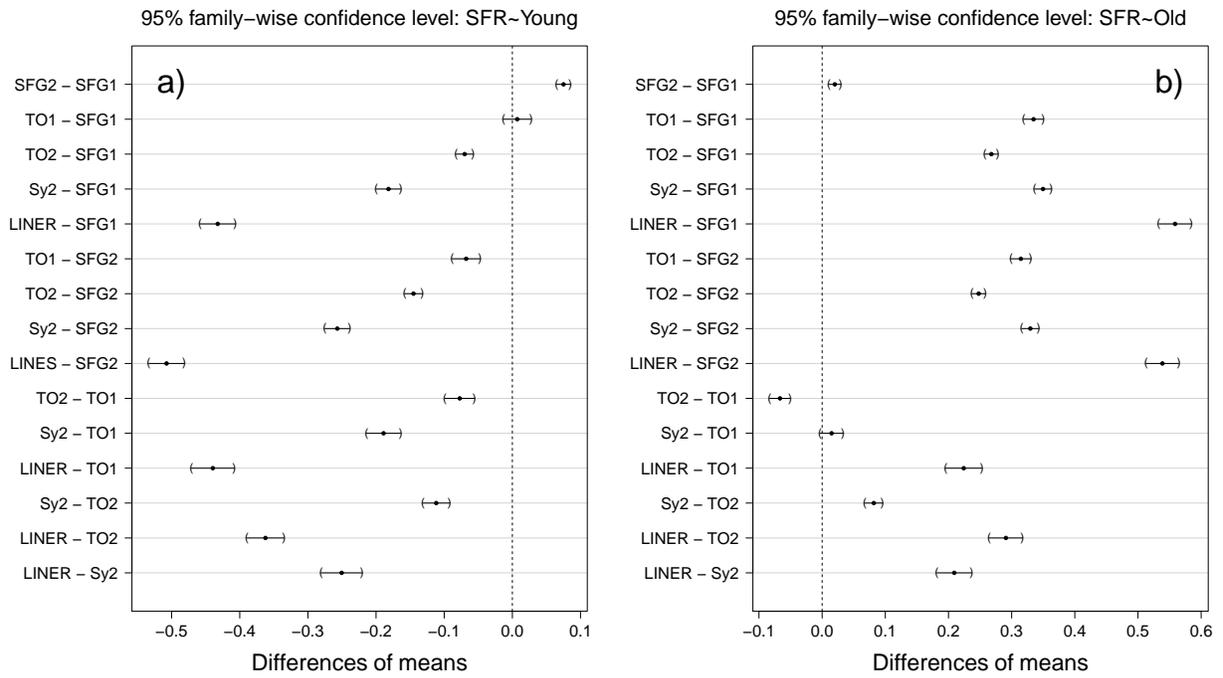} 
\caption{Confidence intervals comparing the two characteristic SFRs in NELGs with
different activity types 
\label{ci_SFR_NELG}} 
\end{figure}

In Fig.~\ref{sfh_NELG}, we show the SFHs of the NELGs in our sample. We find that the SFH varies systematically with the
activity type: the LINERs and Sy2s show their highest SFRs in the past, star formation in the TOs is almost as high
today as in the past, while it is highest at the present time in the SFGs. To make these differences even clearer, we have
extracted from the SFHs two characteristic SFRs as defined in Fig.~6 of \citet{TP13}: SFR$_{\rm Young}$ corresponds to
the maximum SFR of the most recent star formation event, and SFR$_{\rm Old}$ corresponds to the maximum SFR in the past.
The means for the characteristic SFRs in NELGs with different activity types are compiled in Table~\ref{tbl-1}. In
Fig.~\ref{SFR_NELG} we compare their box-whisker plots. In Fig.~\ref{SFR_NELG}a we find that SFR$_{\rm Young}$ increases
along the sequence LINER$\rightarrow$Sy2$\rightarrow$TO2$\rightarrow$TO1$\rightarrow$SFG1$\rightarrow$SFG2. Note that
SFR$_{\rm Young}$ is higher in the SFG2s than in the SFG1s, which is explained by recent bursts of star formation in
SFG2s \citep{TP12b}. In Fig.~\ref{SFR_NELG}b we find that SFR$_{\rm Old}$ decreases monotonically along the sequence
LINER$\rightarrow$Sy2$\rightarrow$TO2$\rightarrow$TO1$\rightarrow$SFG2$\rightarrow$SFG1. The differences between the
SFRs as observed in the box-whisker plots are all confirmed statistically in Fig.~\ref{ci_SFR_NELG}, showing the
confidence intervals for the max-t tests.

\begin{figure} 
\epsscale{0.8} 
\plotone{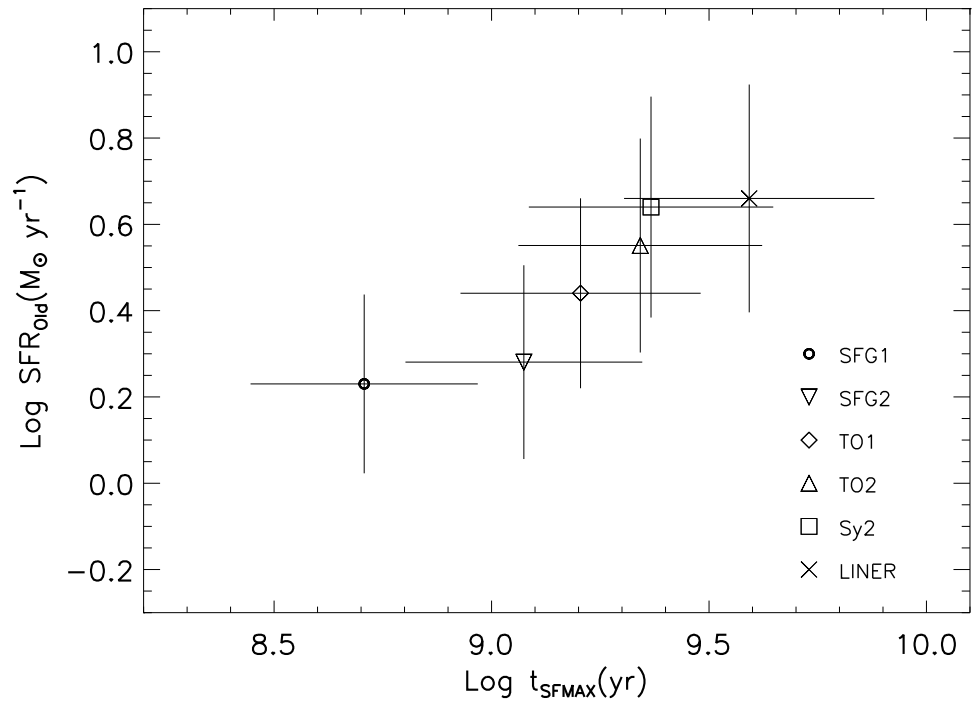} 
\caption{Relation between the maximum star formation in the distant past, and the age of the stars forming this maximum. 
\label{SFROvstSFMAX}} 
\end{figure}

In Table~\ref{tbl-1} we also give the age of the stellar populations producing the maximum SFR in the past, which is consistent with the time in the past when the star formation activity was at its maximum, $t_{\rm
SFMAX}$. In Fig.~\ref{SFROvstSFMAX}, we find that $t_{\rm SFMAX}$ increases in parallel with SFR$_{\rm Old}$. From this
figure we also infer that, although SFR$_{\rm Old}$ are comparable in the LINERs and the Sy2s, the former formed the
bulk of their stars earlier than the latter. Similarly in the SFG2s, despite their recent burst of star formation, the
star formation in the past reaches a maximum long before than in the SFG1s, implying that the stellar populations in the
SFG2s must be older, on average, than in the SFG1s.

\begin{figure} 
\epsscale{0.7} 
\plotone{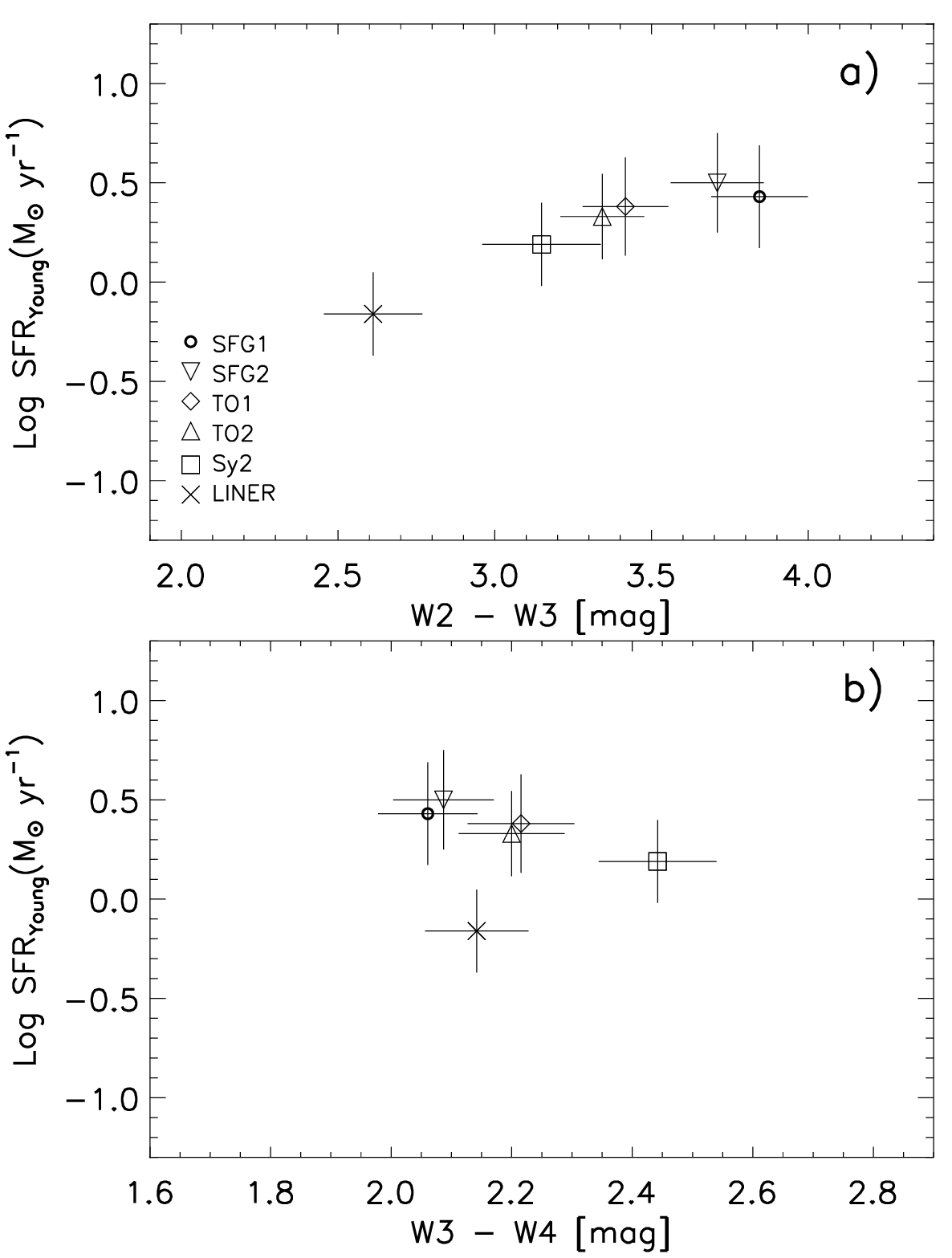} 
\caption{Variations of SFR$_{\rm Young}$ as a function of the MIR colors
of the NELGs with different activity types: a) ${\rm W}2-{\rm W}3$, b) ${\rm W}3-{\rm W}4$.  
\label{SFRYvsMIRcolors}} \end{figure}

\begin{figure} 
\epsscale{0.7} 
\plotone{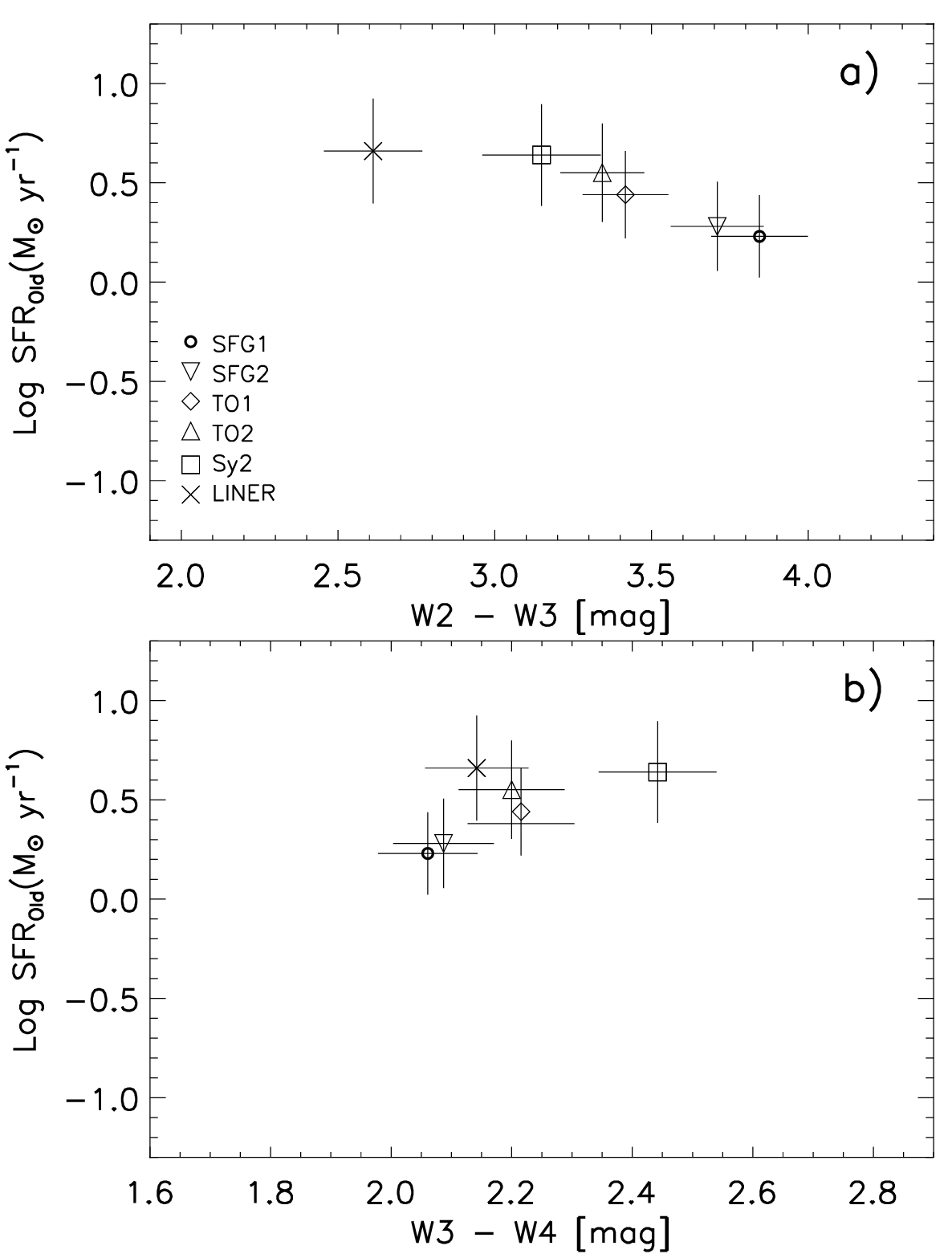} 
\caption{Variations of SFR$_{\rm Old}$ as a function of the MIR colors
of the NELGs with different activity types: a) ${\rm W}2-{\rm W}3$, b) ${\rm W}3-{\rm W}4$. 
\label{SFROvsMIRcolors}} 
\end{figure}

In Fig.~\ref{SFRYvsMIRcolors} and Fig.~\ref{SFROvsMIRcolors} we now compare the two characteristics SFRs with the two
MIR colors used in the MIRDD. In Fig.~\ref{SFRYvsMIRcolors}a, except for the SFG1s and SFG2s as noted before, we find
that the ${\rm W}2-{\rm W}3$ color is indeed well correlated with SFR$_{\rm Young}$. The ${\rm W}2-{\rm W}3$ color of
NELGs are getting redder by $\sim 1.2$ mag as the present level of star formation increases in their host galaxies. In
Fig.~\ref{SFROvsMIRcolors}a, we see the inverse sequence, ${\rm W}2-{\rm W}3$ getting redder as SFR$_{\rm Old}$
increases. On the other hand, in Fig.~\ref{SFRYvsMIRcolors}b and in Fig.~\ref{SFROvsMIRcolors}b, we find no such
correlations: except for the Sy2s, the ${\rm W}3-{\rm W}4$ of NELGs are very similar.

The above comparisons confirm the high sensitivity of the ${\rm W}2-{\rm W}3$ color to the SFH of the NELGs. The position of a NELG in the MIRDD depends not only on its nature as AGN or SFG, but also on its
SFH, and, consequently, on the present level of star formation in its host galaxy. 

These results suggest that the main reason why the MIRDD (and BPT-VO diagram) allows to separate NELGs with different
activity types is because they have different SFHs, i.e., galaxies with different activity types formed their stars
differently
\citep{Dokuchaev91,HamannFerland92,BoyleTerlevich98,Franceschini99,Haiman04,Menanteau05,LouJiang08,Haiman07,Li07,Bennert08,Fanidakis11,Treister11,Donoso12,TP12a,TP12b,Floyd13,TP13}.

\subsection{SED study for the NELGs} 
\label{SED_NELGs}

In the previous section, we have demonstrated that the ${\rm W}2-{\rm W}3$ colors of the NELGs vary systematically with
the SFHs of their host galaxies. However, the MIRDD also separates AGNs from SFGs, which suggests that the AGN component
may also play some role \citep{Coz14}. To determine what is the importance of the AGN component in the SEDs of
emission-line galaxies with different activity types, we have compared the ${\rm W}2-{\rm W}3$ colors of the NELGs with
the colors produced by MIR SED templates as found in the literature. This study is important, not only because it should
allow us to verify independently our interpretation based on the SFHs, but also because the same analysis can be applied
to BLAGNs, for which a stellar population synthesis analysis is obviously not possible.

\begin{figure} 
\epsscale{0.8} 
\plotone{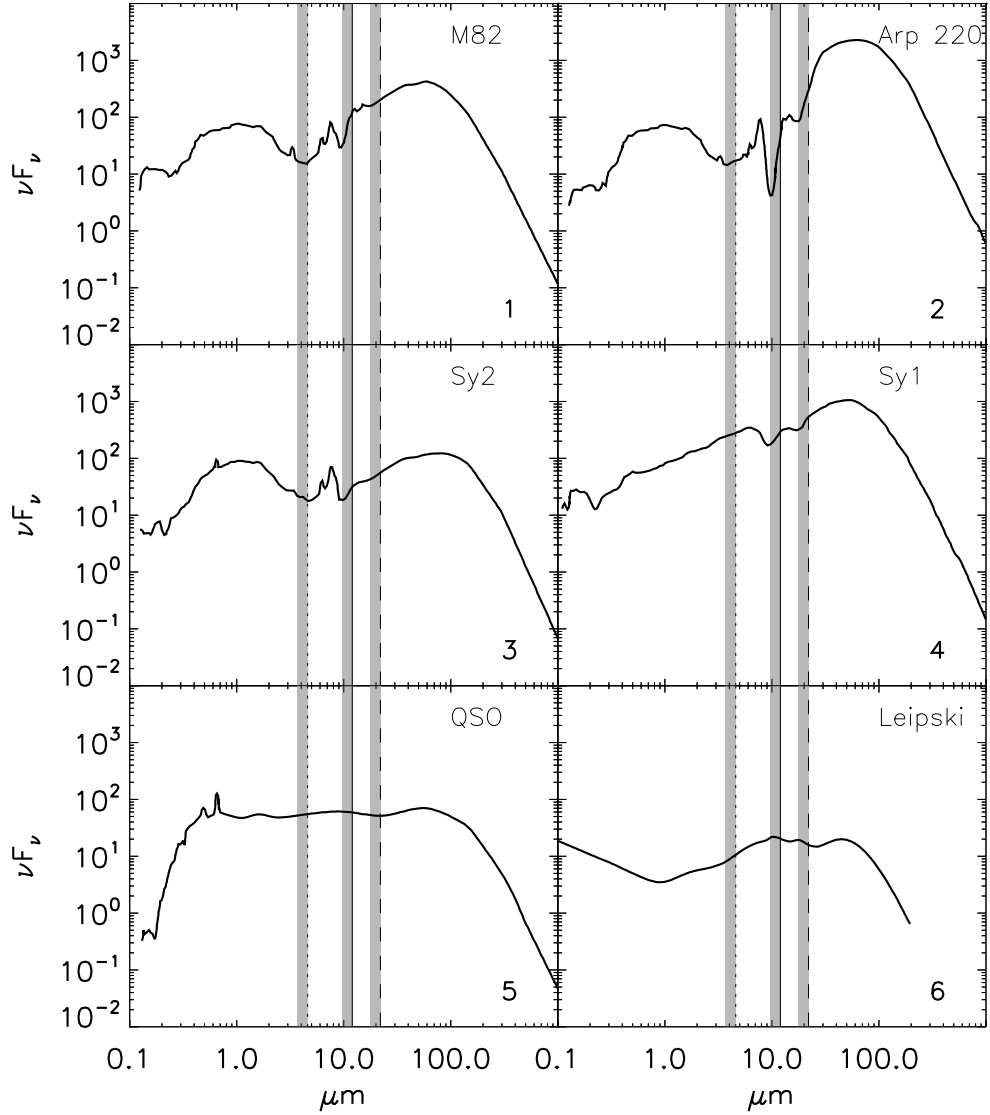} 
\caption{SED templates of NELGs with different activity types. The SEDs are identified as $\#1 =$\ M82, $\#2 =$\ Arp 220, $\#3 =$\
Sy2, $\#4 =$\ Sy1 (Mrk 231), $\#5 =$\ a QSO with relatively mild star formation in its host galaxy, and $\#6 = $ Leipski, a QSO with relatively high star formation in its host galaxy (see explanations in the text). The shaded areas show the regions of the SED that produces the ${\rm W}2-{\rm W}3$ and  ${\rm W}3-{\rm W}4$ colors in the MIRDD from $z=0$ (limit to the red) to $z=0.25$ (limit to the blue). 
\label{SEDtemplates}} 
\end{figure}

For our study we choose six representative SEDs as shown in Fig.~\ref{SEDtemplates}. The first five SEDs come from the
library provided by \citet{Polletta07}. The SEDs \#1 and \#2 were chosen to represent SFGs. These are two typical starburst galaxies, M82 and Arp~220, that due to their high level of star formation are also ultraluminous in infrared. The SEDs~\#3
and \#4 represent, respectivley, a typical Sy2 and Sy1 (Mrk~231). The SED~\#5 was chosen among the different
SEDs proposed by \citet{Polletta07} to represent a typical QSO. Our templates sample in Fig.~\ref{SEDtemplates} is complemented with the SED~\#6, from \citet{Leipski14}, representing a QSO with a strong emission in the FIR.

For normal galaxies (with no dominant AGN component) the interpretation of the SED is straightforward
\citep{Kennicutt03,Murakami07,Wright10,Sheth10,Jarrett13}. Covering the wavelengths from $\sim 1 \micron$ to $100
\micron$, we can distinguish three different regions. Evolved stellar populations, dominated by low mass stars,
produce the light at wavelengths lower than $\lambda \simeq 3 \micron$, forming a bump in emission that culminates at about $1
\micron$. At longer wavelengths than $\lambda \simeq 10 \micron$, there is another emission bump that reaches a maximum at about
$100 \micron$, and which is due to hot dust heated by young, massive stars in active regions of star formation. Finally,
in between these two regions, $3 < \lambda < 10 \micron$, there is a sort of valley, that is also very rich in spectral
structures, showing either unidentified IR bands (UIRBs) in normal galaxies, or PAH emission due to hot stars in starburst-like
galaxies \citep{Xu98,Laurent00,Calzetti07,Hanami12}.  Consequently, the infrared 
SED of normal galaxies is sensitive to stellar light from the evolved stellar populations, as well as to low-temperature
processes from the ISM, and star formation activity
\citep{Kennicutt03,Calzetti07,Wright10,Sheth10,Mateos12,Donoso12,Rosario13,Jarrett13}.

These different components of the MIR SEDs of galaxies, as described above, can easily be distinguished in
Fig.~\ref{SEDtemplates} for emission-line galaxies with different activity types. For example, in SEDs \# 1 and \#2 we clearly see the two bumps at short and long wavelengths and the valley in between. Note that the depth of the valley is more or less pronounced, depending
on the strength of the PAH emission  (obvious in M82, but also in SED~\# 3 for the Sy2) and the presence of UIRBs, which are common features produced by different stellar populations in star forming galaxies. On the other hand, in the SED of AGNs, like SED~\#3 (Sy2) and SED~\#4 (Sy1), the MIR valley more or less disappears. This is due to a high temperature dust component, which is growing in intensity between 1~$\micron$ and 10~$\micron$, and is interpreted to be related to dusty structures, near or at their center, heated by the SMBH  \citep[e.g.,][]{Laurent00,Hanami12,Roseboom13,Leipski14}. In our templates, these hot structures look more prominent in the Sy1 than in the Sy2. However, in both SEDs, we can still see a peak at $100\ \micron$, which implies that star formation may still be active in these galaxies.  The fact that star formation may be present in these AGNs is not unexpected, knowing that Sy2s and Sy1s reside mostly in early-type spiral galaxies. 

The question of determining what is the typical level of star formation in BLAGNs is a difficult one, which is still
open to interpretation \citep[e.g.,][]{Roseboom13}. In QSOs, the problem is even more complicated, because we know very
little about the morphology of their host galaxies, and, consequently, we do not know what is their SFHs. Accordingly,
we have chosen two different SEDs from the literature that represent the extremes. The SED~\#5 represents a QSO where
the AGN component is predominant over the star formation component, while the SED~\#6 represent a QSO, also with a
strong AGN component, but where the level of star formation is comparable to what is observed in ULIRGs, like M82 or Arp
220. \citet{Leipski14} determined this SED from a sample of QSOs at high redshifts, concluding that active star formation
is a common feature of AGNs at high redshifts \citep[see also][]{Bertoldi03,Wang08,Wang10}.

To determine which SED in our sample of templates better reproduces the ${\rm W}2-{\rm W}3$ colors of NELGs with
different activity types, we have calculated the colors expected from these SEDs at different redshifts. It
is well known that due to the increase in redshift, the wavelengths that enter a set of filters correspond to a bluer
part of the SED. This effect is at the basis of the K-correction. In Fig.~\ref{SEDtemplates}, the shaded areas show the
regions of the SEDs that produce the ${\rm W}2$, ${\rm W}3$, and ${\rm W}4$ magnitudes, from $z=0$ (limit to the red) to
$z=0.25$ (limit to the blue). We have verified that due to the different filter responses, the difference in width of
the filters (especially W3) does not produce any significant difference in colors. Therefore, to calculate the different
magnitudes, we have used the fluxes at the center wavelengths of the filters at different redshifts (the shaded areas), assuming filter responses of one. To convert the fluxes to magnitudes, we have applied the flux correction factors for a power law spectrum with spectral index $\alpha = 0$, which, according to \citet{Wright10}, is the most general case.

\begin{figure} 
\epsscale{0.65} 
\plotone{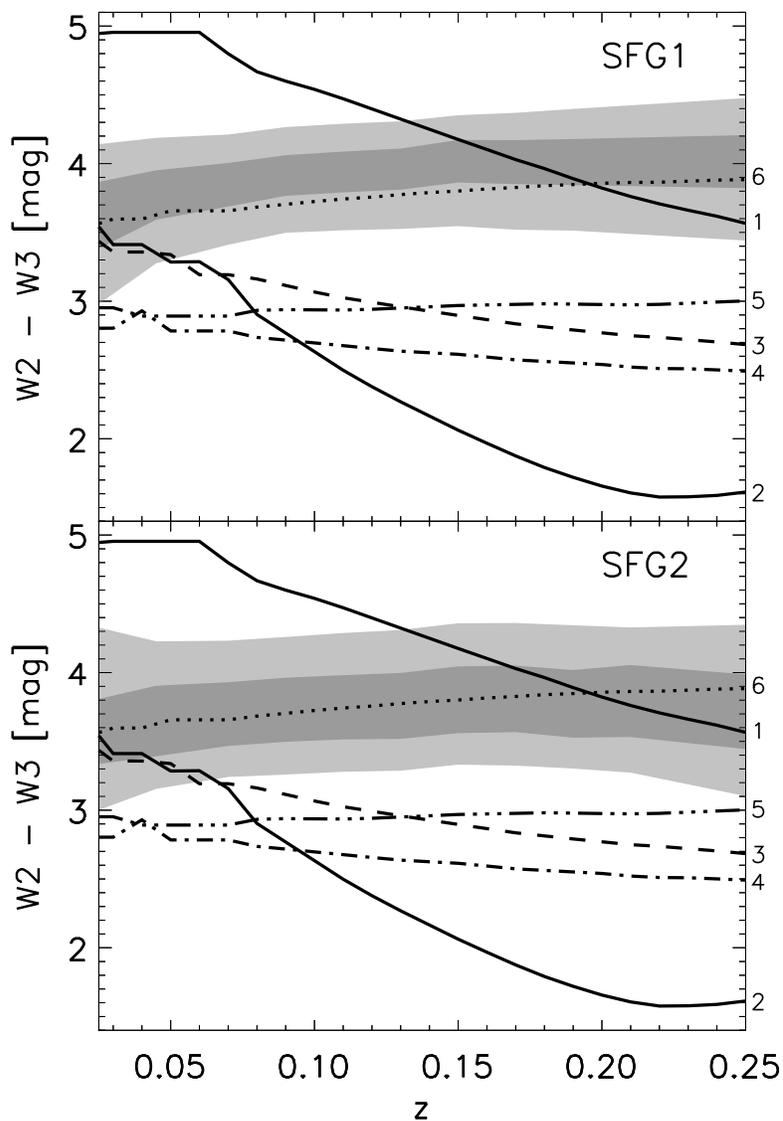} 
\caption{Observed variations of the ${\rm W}2-{\rm W}3$ colors of the SFGs from $z = 0$ to $z = 0.25$. The light gray area corresponds to the 5\% and 95\% percentiles, whereas the darker area corresponds to the 25\% and 75\% percentiles (or first and third quartiles). The different curves, identified by the numbers in the right margin, correspond to the colors at different redshifts, as predicted by the SEDs with the same number in Fig.~\ref{SEDtemplates}.  
\label{SED_SFG}} 
\end{figure}

\begin{figure} 
\epsscale{0.7} 
\plotone{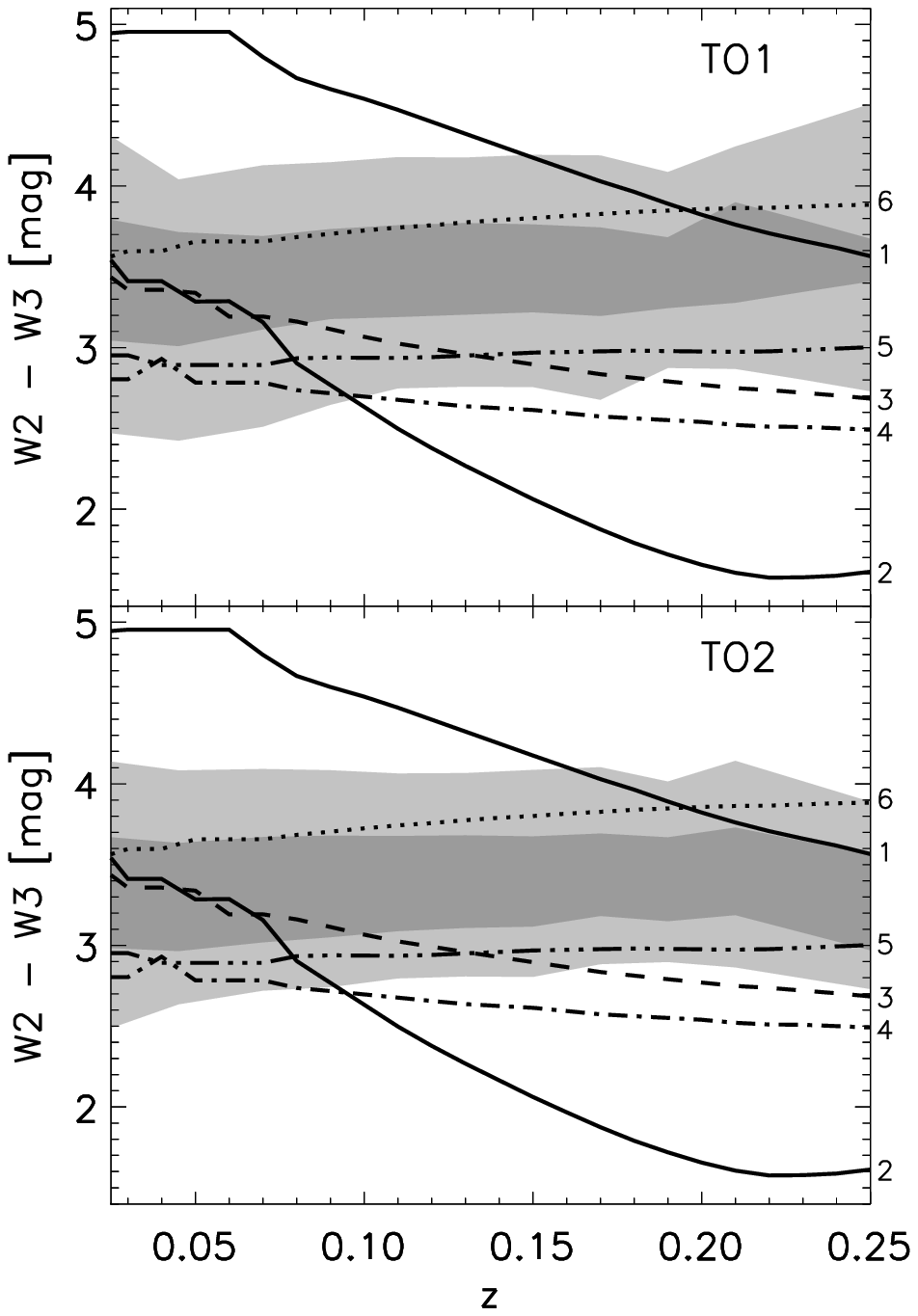} 
\caption{Observed variations of ${\rm W}2-{\rm W}3$ color of the TOs from $z = 0$ to $z = 0.25$. The shaded areas and curves are as explained in Fig.~\ref{SED_SFG}.
\label{SED_TO}}
\end{figure}

\begin{figure} 
\epsscale{0.7} 
\plotone{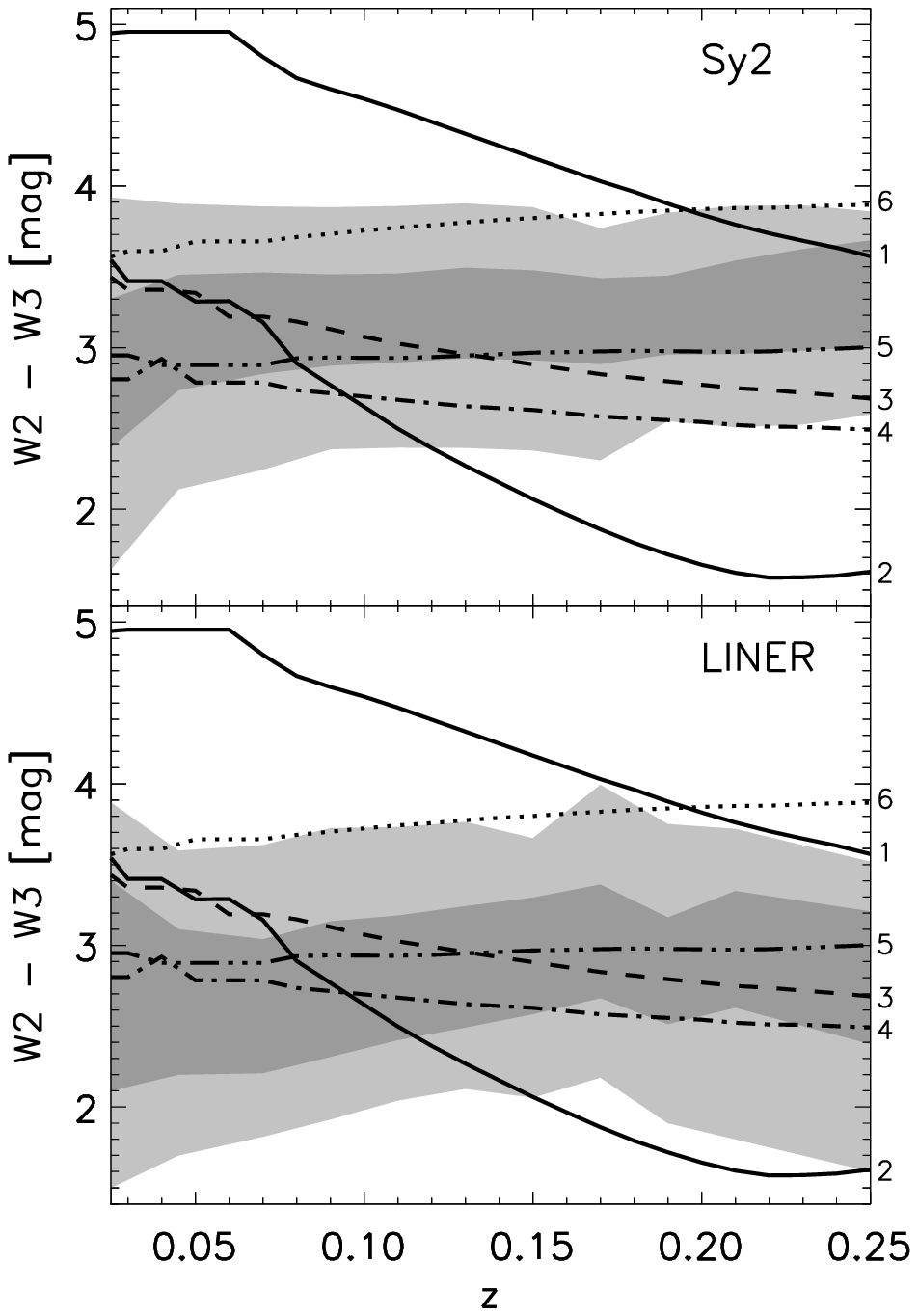} 
\caption{Observed variations of ${\rm W}2-{\rm W}3$ color of the Sy2s and LINERs from $z = 0$ to $z = 0.25$. The shaded areas and curves are as explained in Fig.~\ref{SED_SFG}.
\label{SED_AGN}} 
\end{figure}

We show the variations of ${\rm W}2-{\rm W}3$ colors between $z = 0$ and $z = 0.25$ for the SFGs in Fig.~\ref{SED_SFG},
for the TOs in Fig.~\ref{SED_TO}, and for the AGNs in Fig.~\ref{SED_AGN}. In all these figures the light grey areas
correspond to the 5\% and 95\% color percentiles, whereas the darker areas correspond to the 25\% and 75\% percentiles.
Note that the dispersions of the ${\rm W}2-{\rm W}3$ colors in all these figures are remarkably small, the darker areas
having typical widths $\sim 0.5-0.7$ mag. Therefore, we identify the ``typical'' SEDs of NELGs with different activity
types with the SEDs that best reproduce their median ${\rm W}2-{\rm W}3$ colors at redshift $z = 0$. The K-correction,
consequently, would correspond to the color correction implied by this SED at different redshifts.

For the SFG1s and SFG2s, in Fig.~\ref{SED_SFG}, the best SEDs at $z=0$ are the SED~\#2 (Arp~220), for a typical
starburst galaxy, the SED~\#3 for a typical Sy2 and the SED~\#6, which is the SED of a QSO hosted by a galaxy
experiencing a high level of star formation. All these SEDs have in common that they imply a high level of
star formation in the host galaxies, which suggests that this characteristic is a necessary condition to explain the
position occupied by the SFGs in the MIRDD.

For the TO1s, in Fig.~\ref{SED_TO}, the same three SEDs as for the SFGs fit their median color at $z=0$, suggesting that
their host galaxies are also actively forming stars. However, the medians of these SEDs are slightly too high to fit the
median of the TO2s, and a SED giving a value in between SED~\#3 and SED~\#5 would be better. This would be consistent
with a lower level of star formation in the TO2s as compared to that of the TO1s and SFGs.

Similarly, for the Sy2s and LINERs in Fig.~\ref{SED_AGN}, we see a clear shift in median colors toward the blue. The
best SEDs at $z=0$ for the Sy2s are either the SED \#4, which is the SED of Mark 231, a Sy1 galaxy, or SED \#5, which is
the SED of a typical QSO with a low star formation component. For the LINERs both SEDs, SED \#4 or SED \#5, would only
fit the reddest of these galaxies, their median being slightly bluer. This is consistent with lower star formation in
LINERs than Sy2s.

From the identification of the best SEDs at $z=0$, we conclude that the different ${\rm W}2-{\rm W}3$ colors of the
NELGs with different activity types can be explained by varying the contributions of the AGN and star formation
components in their SEDs. The AGN component is more prominent in the LINERs and Sy2s, whereas the star formation
component is more important in the TO1s and SFGs, while the TO2s are between the Sy2s and TO1s. This result is in good
agreement with our analysis of the SFHs of these galaxies.

But what about the variations of the ${\rm W}2-{\rm W}3$ colors with the redshift? Examining Fig.~\ref{SED_SFG},
Fig.~\ref{SED_TO}, and Fig.~\ref{SED_AGN}, the most important characteristics of the NELGs in our sample is that they
all show almost constant MIR colors from $z = 0$ up to $z = 0.25$. This explains why NELGs with different activity types
form specific regions in the MIRDD. However, from the point of view of the SED, we can only reproduce this constancy in
colors if it includes an AGN component. Obviously, for the Sy2s and LINERs in Fig.~\ref{SED_AGN} this does not
pose any problem. Consequently, adopting SED~\#4 the K-correction for the colors of the Sy2s at $z = 0.25$ would be 0.6 mag to the blue,  whereas only $\sim 0.3$\ mag would be needed adopting SED~\#5. Similarly for the LINERs at $z = 0.25$, a
K-correction in color of 0.3 mag to the blue would be needed adopting SED~\#4, while a K-correction of $\sim 0.2$\ mag to the red would be needed adopting SED~\#5. In both cases, the best fit with lower K-correction is SED~\#5. However, considering the colors of  SED~\#6, it seems clear that a QSO SED with slightly higher star formation would fit better the Sy2s, while a QSO SED with lower star formation would fit the LINERs. 
 
On the other hand, in Fig.~\ref{SED_TO}, adopting for the TOs the SED~\#2 for a starburst would imply increasingly large
K-corrections to the blue for galaxies at $z > 0.07$, reaching $\sim 2.0$ mag at $z = 0.25$ for theTO1s, and slightly
less for the TO2s. However, this choice of SED for the TOs is not unique, since, according to the BPT-VO diagram, TOs
are also assumed to have an AGN at their center. For example, adopting the SED~\#3 would reduce the K-correction to the
blue to $\sim 0.8$ mag at $z = 0.25$\ for the TO1s, and slightly less for the TO2s. However, the SED that predicts the
lowest K-correction for the TOs is SED~\#6, producing 0.3 mag to the red for the TO1s and 0.5 mag for the TO2s. This is
the SED of a QSO with strong star formation. Therefore, a solution consistent with the average of SED~\#5 and SED~\#6,
would neatly fit the median colors of the TOs at any redshifts, with a minimal K-correction up to $z = 0.25$. But, this
would imply increasing the level of star formation in TOs at higher redshifts.

Obviously, the problem of keeping a constant ${\rm W}2-{\rm W}3$ color as the redshift increases is much more
critical for the SFGs in Fig.~\ref{SED_SFG}. Coincidentally, adopting the SED~\#6 for the SFGs would predict almost no
K-correction up to $z = 0.25$ for the SFG2s, and a slightly blue correction for the SFG1s. But, this solution  
implies an AGN component is needed in the SEDs of these galaxies. Note that this might explain why the colors of the SFGs in the MIRDD do not fit those of a black body. Alternatively, a solution based on an average of SED~\#1 with SED~\#2, with a bias toward low star formation (SED\#2) at low redshift and a bias toward higher star formation (SED\#1) at higher redshift, may reproduce the constant MIR colors of
the SFGs without the need for a high K-correction. Adopting this solution, star formation in SFGs would then be expected to increase to the level of ULIRGs at higher redshifts \citep[][]{TP12b}\footnote{Alternatively, in \citep{Coz14} it was suggested that a SED with multiple black body components could produce MIR colors consistent with a power law. For the SFGs, such a SED could be consistent with a sequence of bursts of star formation \citep{Coz96,Coz01,TP12b}.}.   

According to our analysis of the SEDs, we conclude that it is possible to reproduce the ${\rm W}2-{\rm W}3$ colors, and
the lack of variation with the redshift from $z = 0$\ to $z = 0.25$ with minimal K-corrections, by assuming that: 1) an
AGN component dominates the SEDs of LINERs and Sy2s, the Sy2s showing a higher level of star formation than the LINERs,
2) the SEDs of TOs must include both, an AGN and star forming components, with a level of star formation decreasing
along the sequence TO1$\rightarrow$TO2$\rightarrow$Sy2, and 3) either the SFGs show an AGN with high level of star
formation, or star formation in these galaxies gradually increases to the level of ULIRGs at higher redshift. In
general, these results are in good agreement with our analysis of the SFHs, showing that the ${\rm W}2-{\rm W}3$ colors
of NELGs with different activity types follow an increase in star formation in their host galaxies.

\subsection{Comparing BLAGNs with NELGs in the MIRDD} 
\label{MIRDD_BLAGN}

\begin{figure} 
\epsscale{0.6} 
\plotone{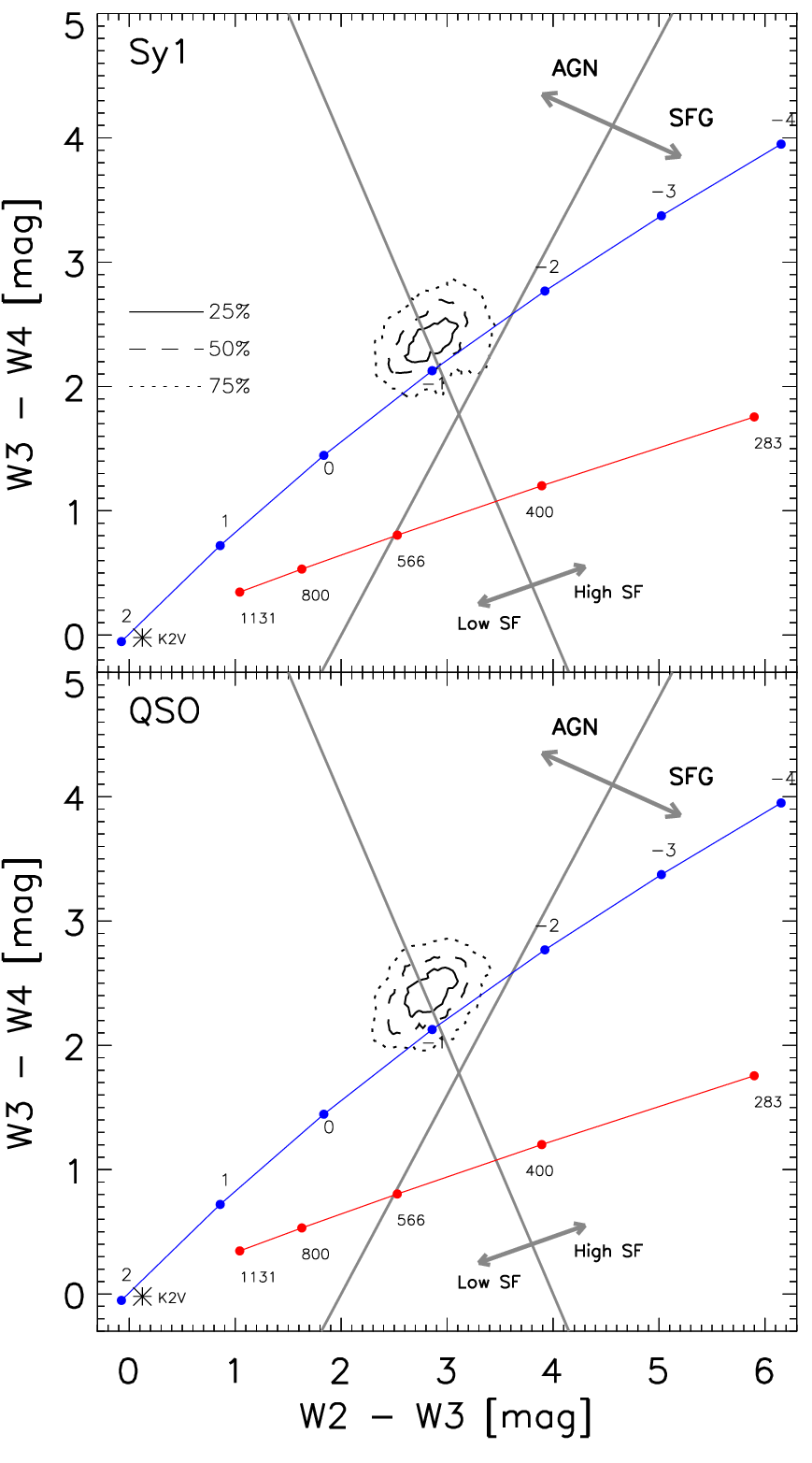} 
\caption{MIRDD for the nearby BLAGNs. In a) the Sy1s and in b) the QSOs. The boundaries and curves are as described in Fig.~2. 
\label{mirdd_BLAGN}}
\end{figure}

\begin{figure} 
\epsscale{0.8} 
\plotone{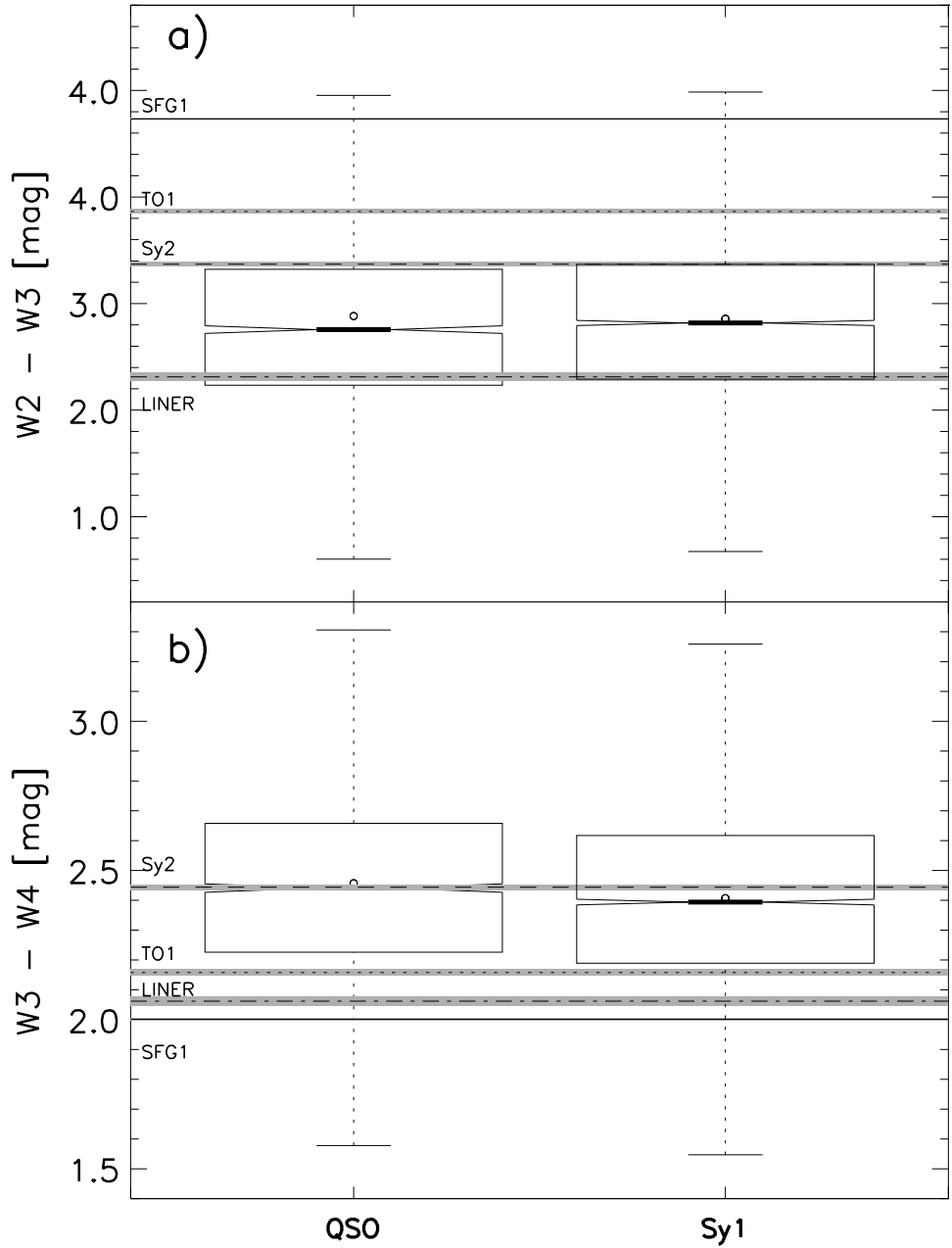} 
\caption{Box-whisker plots comparing the distributions of MIR colors of the nearby BLAGNs with the median colors (and notches) of the Sy2s, the LINERS, the TO1s and SFG1s.
\label{bw_BLAGN}} 
\end{figure}

\begin{figure} 
\epsscale{1} 
\plotone{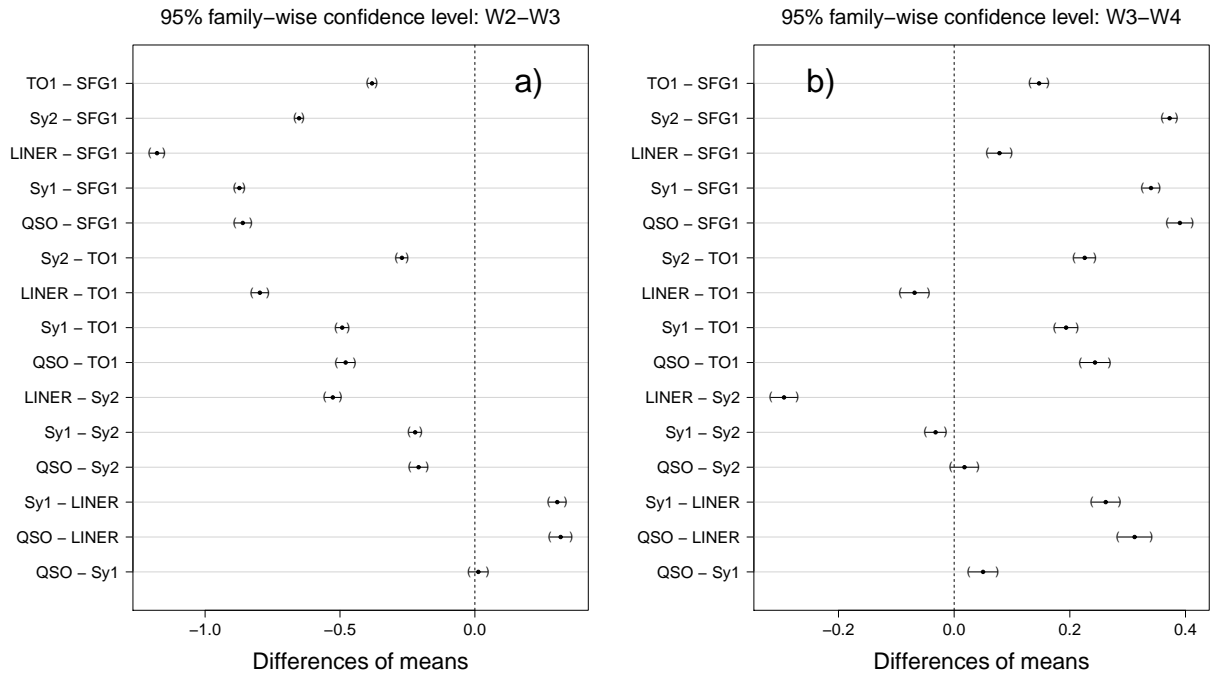} 
\caption{Confidence intervals for the comparison of MIR colors of BLAGNs with the colors of the Sy2s, the LINERS, the TO1s and SFG1s: a) ${\rm W}2-{\rm W}3$ colors, b) ${\rm W}3-{\rm W}4$ colors. \label{ci_BLAGN}} 
\end{figure}

In Fig.~\ref{mirdd_BLAGN}, we present the MIRDD for the BLAGNs. We find no difference between the positions of the Sy1s
and QSOs.  They both occupy the same region of the MIRDD, which is intermediate between the positions of the LINERs and
Sy2s. This is more obvious in Fig.~\ref{bw_BLAGN} where we present the box-whisker plots, comparing the colors of the
nearby BLAGNs with the median values observed for the NELGs with different activity types. The similarities between the nearby
QSOs and Sy1s are confirmed statistically by the confidence intervals presented in Fig.~\ref{ci_BLAGN}: 1) the Sy1s and QSOs have similar ${\rm W}2-{\rm W}3$ colors, whereas the QSOs have slightly redder
${\rm W}3-{\rm W}4$ colors than the Sy1s, 2) the BLAGNs have redder ${\rm W}2-{\rm W}3$ colors than the LINERs, and are
bluer than the Sy2s by about the same amount ($\sim 0.2$ mag), and 3) the ${\rm W}3-{\rm W}4$ colors of the BLAGNs are
comparable to those of the Sy2s, being much redder than the LINERs.

\begin{figure} 
\epsscale{0.8} 
\plotone{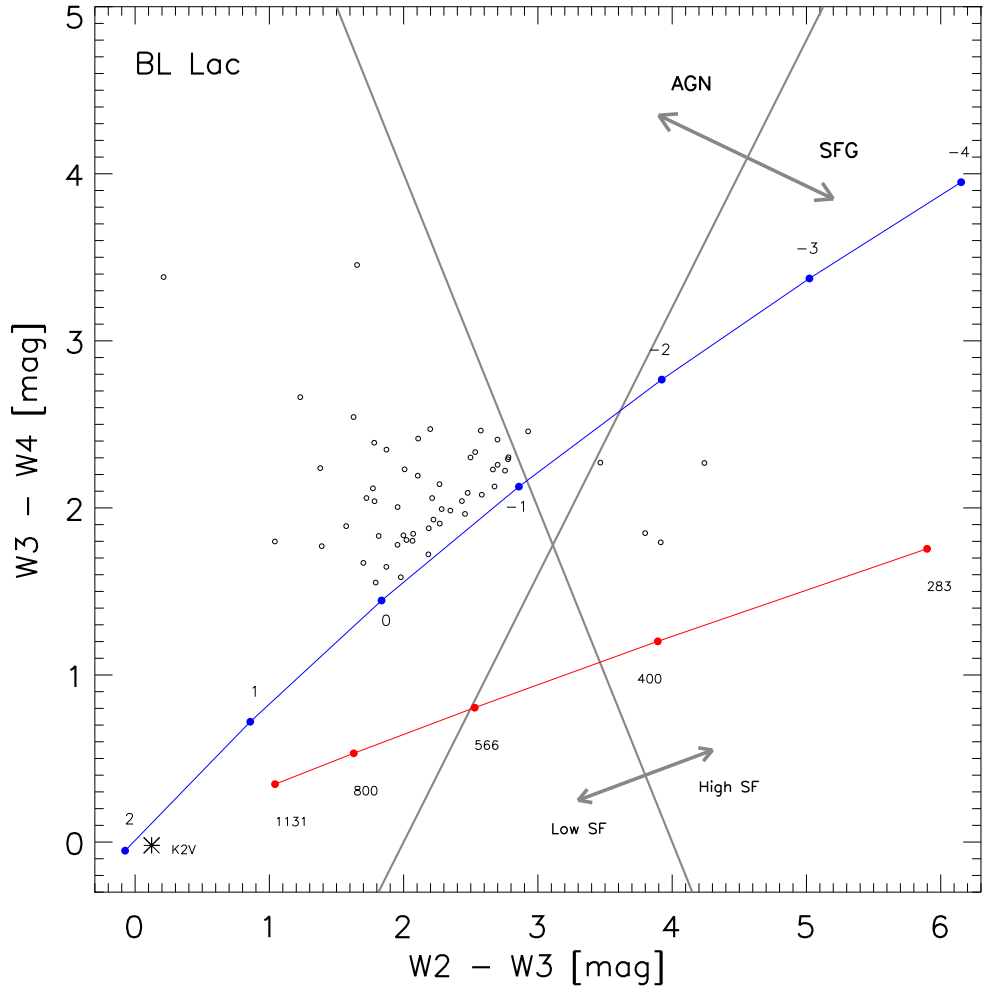} 
\caption{MIRDD for the nearby BL~Lac objects. The boundaries and curves are as described in Fig.~2. 
\label{mirdd_BLLAC}}
\end{figure}

\begin{figure} 
\epsscale{0.7} 
\plotone{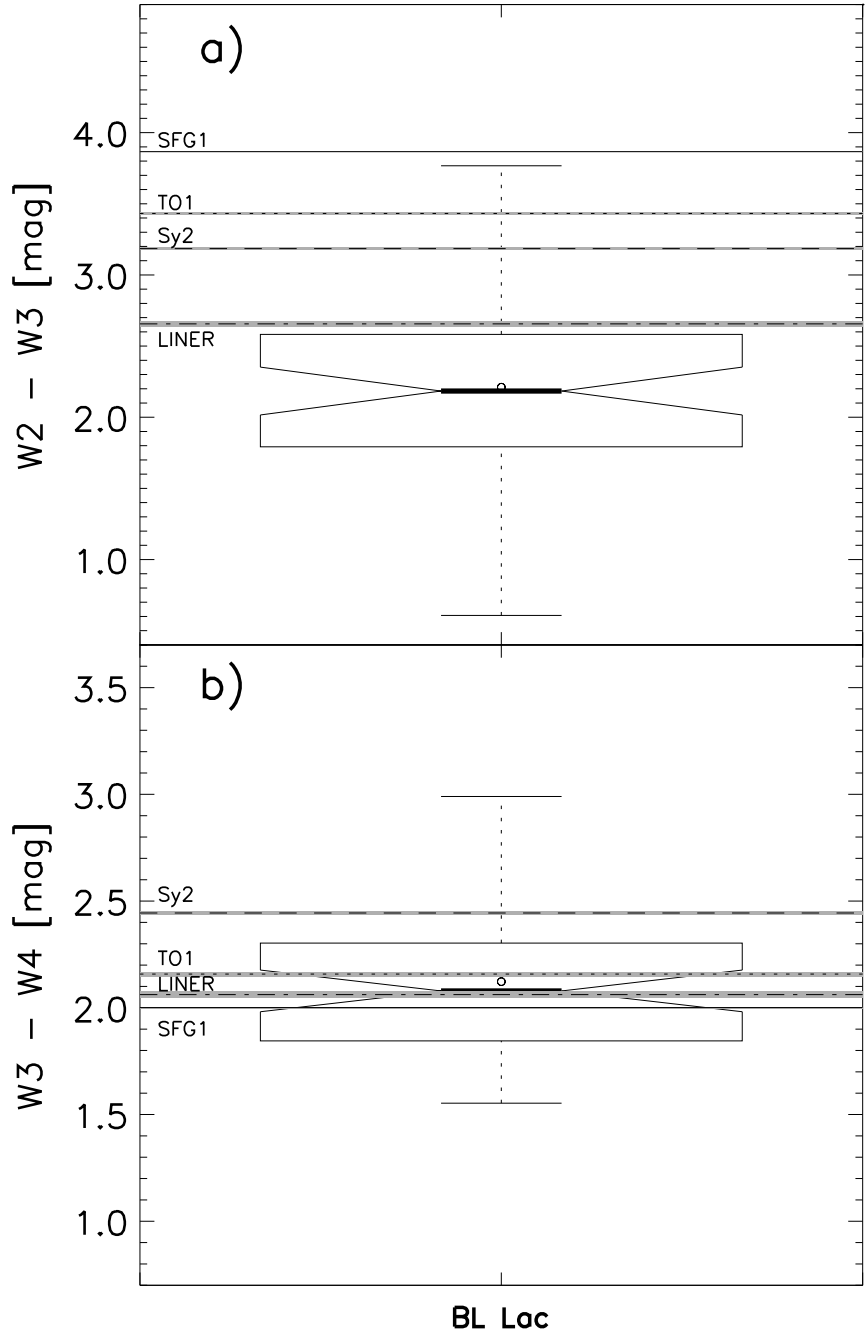} 
\caption{Box-whisker plots comparing the distributions of MIR colors of the nearby BL~Lac objects with the median colors of the Sy2s, the LINERS, the TO1s and SFG1s.
\label{bw_BLLAC}} 
\end{figure}

\begin{figure} 
\epsscale{1} 
\plotone{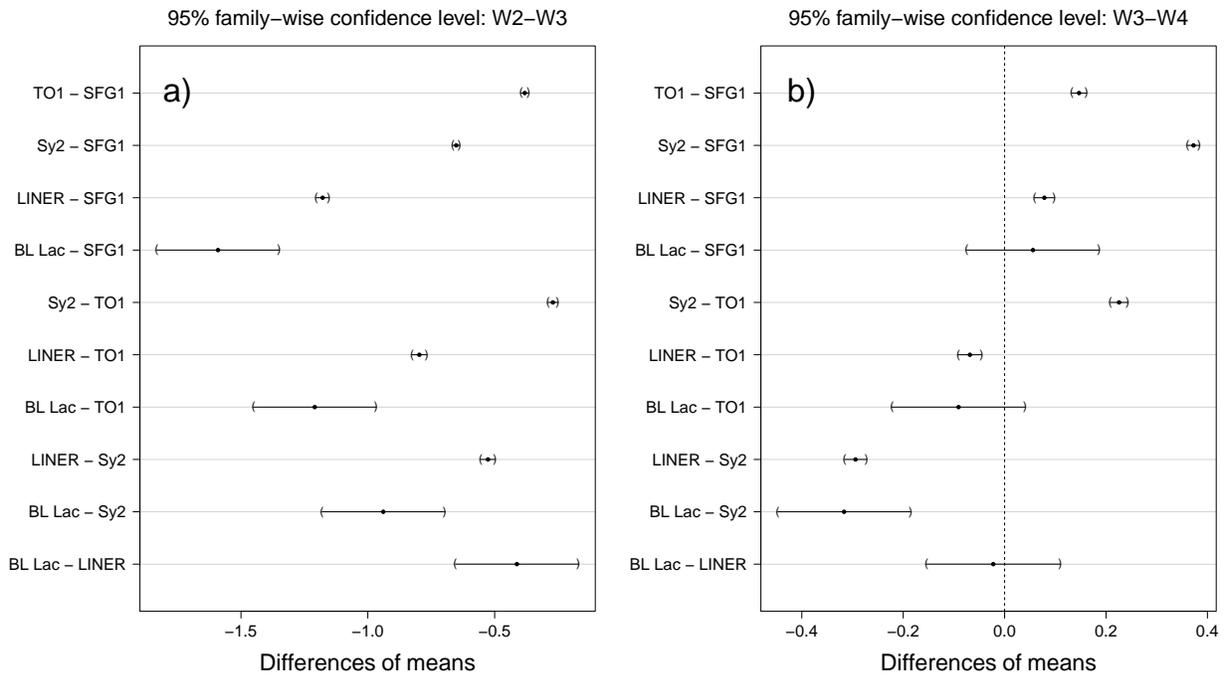} 
\caption{Confidence intervals for the comparison of the MIR colors of the nearby
BL~Lac objects with the colors of the Sy2s, the LINERS, the TO1s and SFG1s. 
\label{ci_BLLAC}} 
\end{figure}

We present the MIRDD for the BL Lac objects in Fig.~\ref{mirdd_BLLAC}. Their positions follow a power-law with spectral index
$\alpha$ varying between 0 and $-1.5$, exactly like the LINERs. The similarity in MIR colors between the BL Lac objects and
LINERs can also be observed in Fig.~\ref{bw_BLLAC}, where we show the box-whisker plots for the MIR colors: the BL Lac
are bluer by $\sim 0.4$ mag  in ${\rm W}2-{\rm W}3$  than the LINERs (only $\sim 0.2$ if we consider the BEff using $w?Gmag$), but showing similar ${\rm W}3-{\rm W}4$ colors. The similarities between the BL Lac and LINERs are confirmed by the confidence intervals from the max-t
tests as presented in Fig.~\ref{ci_BLLAC}. The fact that the LINERs and BL Lac objects have comparable MIR colors is important, since these two galaxies constitute the two extremes in terms of AGN luminosity.  

Taken as a whole, one can see that at low redshift ($z \le 0.25$) all the different AGNs known, namely, the LINERs, the
Sy2s and Sy1s, the QSOs and BL~Lac, occupy the same region of the MIRDD, their colors
tracing a continuous sequence that is consistent with the color sequence produced by a power law, $F_\nu = \nu^{\alpha}$, where
the spectral indice, $\alpha$, varies between 0 and $-2$, in good agreement with what is seen in the optical-ultraviolet part of the spectra of 
AGNs with different luminosities. In the MIRDD, one can also observe that the ${\rm W}2-{\rm W}3$ colors of the AGNs
become redder along the sequence BL~Lac$\rightarrow$LINER$\rightarrow$QSO/Sy1$\rightarrow$Sy2. This is different from the AGN luminosity, which decreases along the sequence BL~Lac/QSO$\rightarrow$Sy1$\rightarrow$Sy2$\rightarrow$LINER. This suggests that the sequence in ${\rm W}2-{\rm W}3$ color is not driven by the AGN component. 

\subsection{SED study for the BLAGNs} 
\label{SED_BLAGNs}

\begin{figure} 
\epsscale{0.7} 
\plotone{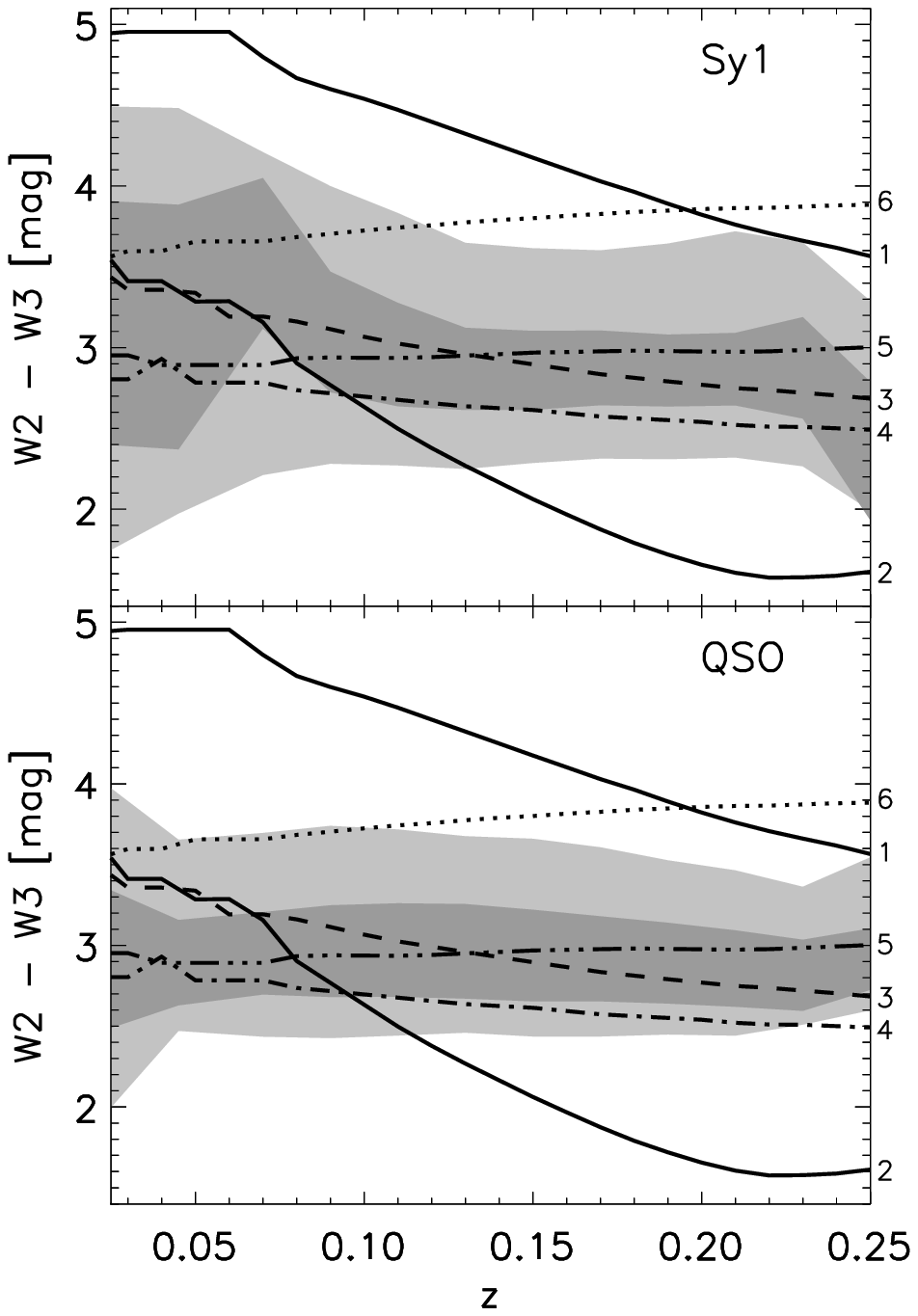} 
\caption{Observed variations of ${\rm W}2-{\rm W}3$ color of the Sy1s and QSOs from $z = 0$ to $z = 0.25$. The shaded areas and curves are as explained in Fig.~\ref{SED_SFG}.
\label{SED_BLAGN}}
\end{figure}

In Fig.~\ref{SED_BLAGN}, we show the variations of the ${\rm W}2-{\rm W}3$ colors of the BLAGNs from $z = 0$ to $z =
0.25$. For the Sy1s, there are no clear best fits at $z = 0$. The reddest Sy1s can be fitted by the SED~\#2, the SED~\#3, or the SED \#6, exactly like for the SFGs, while the bluest one can only be fitted by either SED~\#4 or SED~\#5, like for the Sy2s and LINERs. On the other  hand, only SED~\#3 (for a Sy2) or SED~\#5 would reproduce the colors of Sy1s at any redshift, with minimal K-corrections. These fits seems to imply that the Sy1 host galaxies show a variety of star formation, some being as active as Sy2s, and others much less active, similar to the LINERs. This is consistent with their intermediate positions in the MIRDD.

For the QSOs, the best fits at $z = 0$ are either SED~\#4 or SED~\#5. However, SED~\#5 would predict smaller K-corrections than SED~\#4 over the whole range in redshift.  The differences in fits suggest that at low redshifts star formation is higher in Sy1s than in QSOs, while it is comparable at high redshifts. The trend for the BLAGNs is thus very clear, they both require a strong AGN component, and some level of star formation
in their host galaxies, with possibly a slightly higher star formation level, on average, in the Sy1s than in the QSOs. 

\begin{figure} 
\epsscale{0.7} 
\plotone{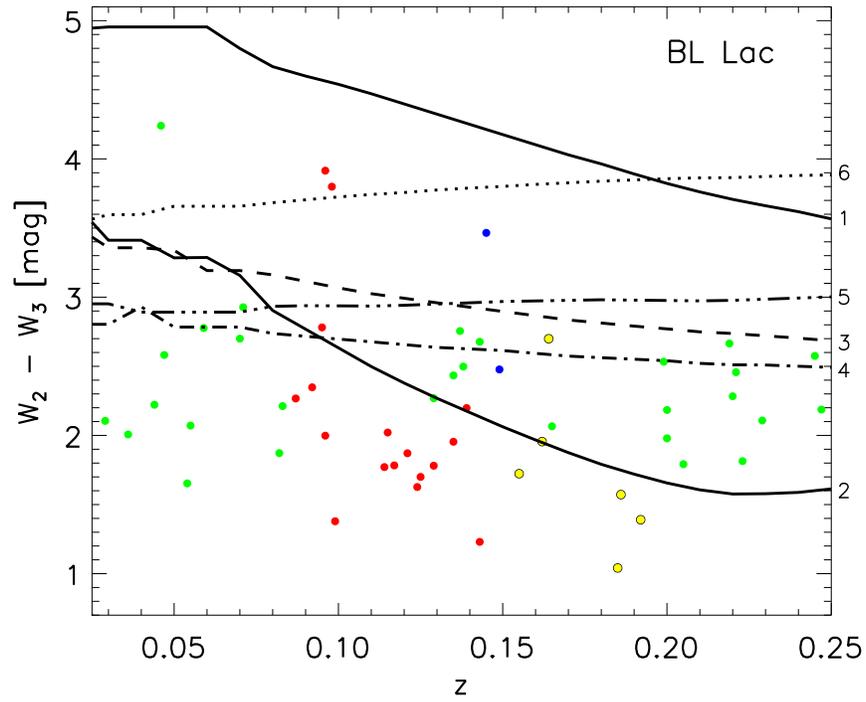} 
\caption{Observed variations of ${\rm W}2-{\rm W}3$ color of the BL Lac objects from $z = 0$ to $z = 0.25$. The colors correspond to the different level of resolutions according to the WISE photometry flag ($ext\_flags$): 0 (green), 1 (red) , 234, (blue), and 5 (yellow).
\label{SED_BL}}
\end{figure}

\begin{figure} 
\epsscale{0.6} 
\plotone{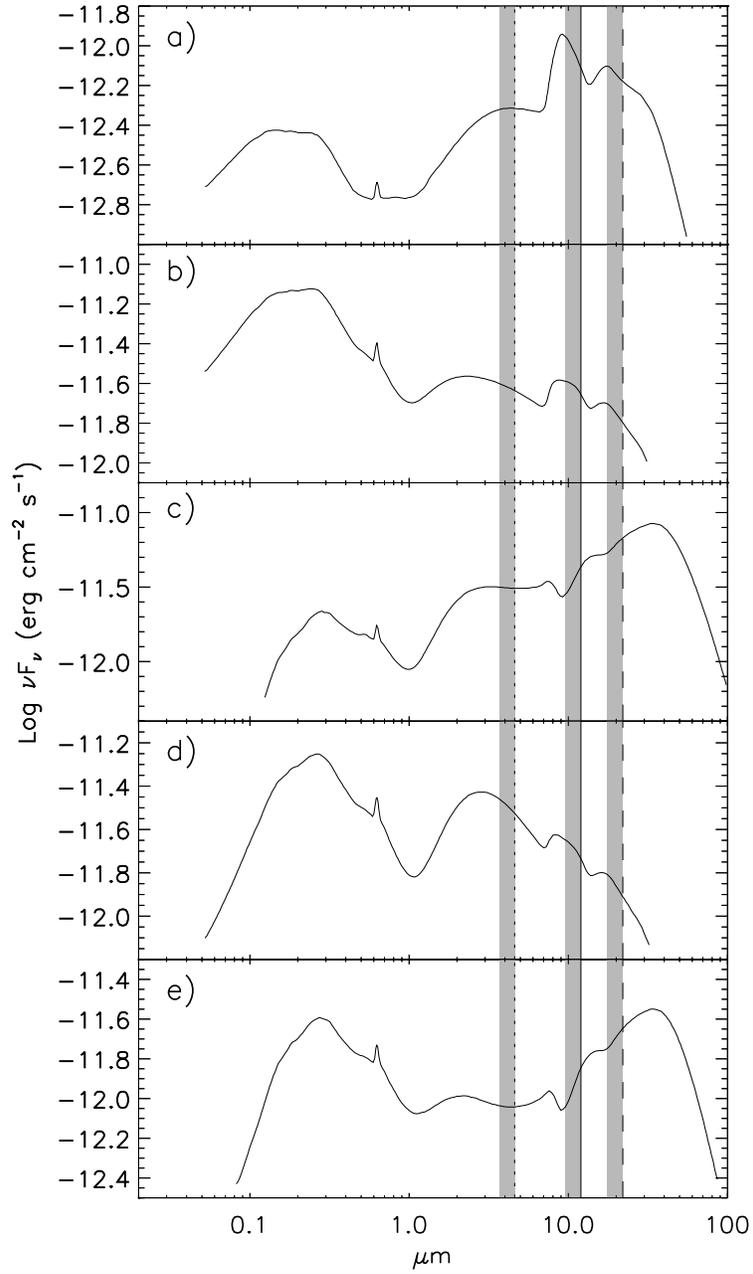} 
\caption{SEDs for QSOs with prominent AGN components, as establsihed by \citet{Roseboom13}.  The shaded areas show the regions of the SED that produce the ${\rm W}2-{\rm W}3$ and ${\rm W}3-{\rm W}4$ colors from $z=0$ to $z=0.25$. 
\label{SEDtemplates2}} 
\end{figure}

\begin{figure} 
\epsscale{0.8} 
\plotone{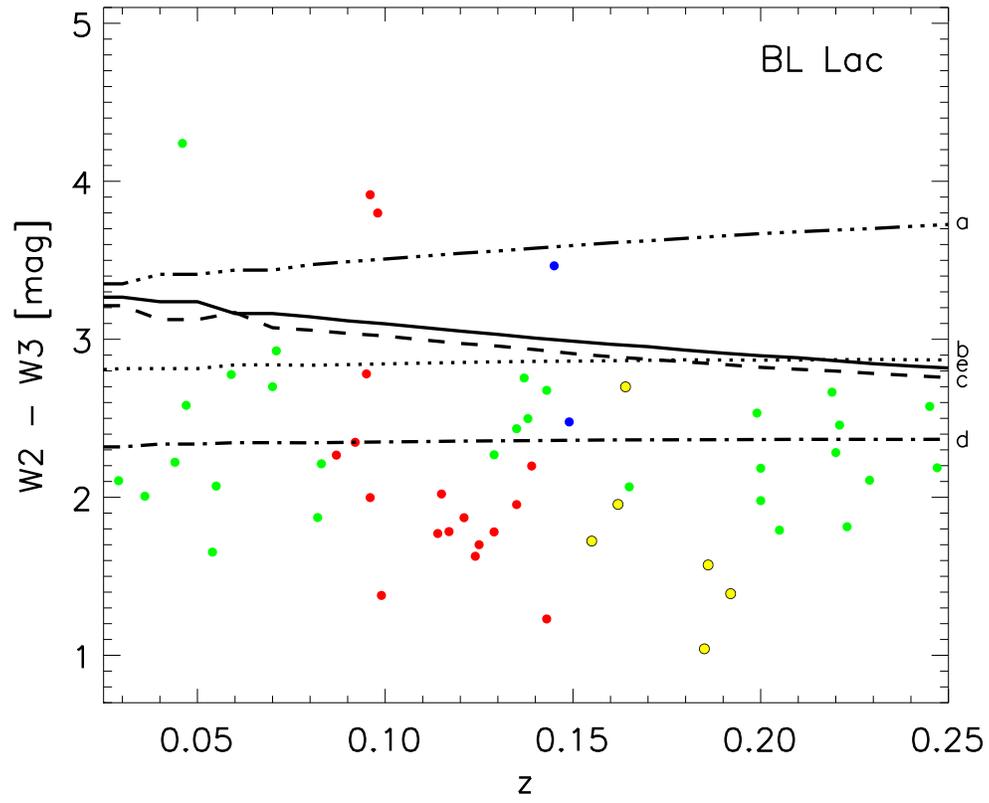} 
\caption{Comparison of the variation of ${\rm W}2-{\rm W}3$ color with redshift for the BL~Lac  using the SEDs in Fig.~\ref{SEDtemplates2}. 
\label{SED2_BL}}
\end{figure}

The case for the BL Lac objects in Fig.~\ref{SED_BL} is also very clear. None of the six SEDs in our templates sample can reproduce their ${\rm W}2-{\rm W}3$ at  $z = 0$, these galaxies being much too blue in MIR. What we need, it seems, would be a SED with a much higher flux at W2 than at W3, which, according to our interpretation, would be consistent with a strong AGN component in a galaxy with very little, or no star formation. To illustrate this point, we show in  
Fig.~\ref{SEDtemplates2} the SEDs observed by \citet{Roseboom13}.  Although the interpretation of these SEDs made by these authors do not agree with ours (their models assume that all the IR emission is due to the AGN), we note that the SEDs identified as b) and d) would have the correct form for the BL Lac objects, showing higher flux at W2 than at W3. Indeed, in Fig.~\ref{SED2_BL} one can see that the SED that best reproduces the colors of the BL Lac objects is SED~d, confirming that for the BL Lac, a strong AGN component is necessary, and, obviously (there are no peaks at 100$\micron$ in these SEDs), no star formation  is expected in their host galaxies. Again, this is consistent with what we know a priori of these objects: BL Lac object are luminous AGNs in elliptical galaxies \citep{Kotilainen98,Scarpa00,Falomo14}. 

\section{Discussion} 
\label{DISCUSSION}

According to the BPT-VO diagram, there are only two main sources of ionization of the gas in NELGs at low redshift ($z
\le 0.25$): young massive stars in star forming regions and a SMBH in AGNs. In many AGNs these two sources
are known to be operating at the same time, which may complicate their identification
\citep{HFS97,Wu98,Maoz99,Vila-Vilaro00,Kew01,Carter01,CZ06,GD08,Yuan10,Carpineti12,Feltre13,Coz14}. So, the question is
can we confirm this simple ``standard'' view in the MIR? Our study based on the SFHs and SEDs suggests that the answer is indubitably yes.
In the MIRDD, NELGs with different activity types, as determined using the BPT-VO diagram, show clear distinctive
colors, due to the variation of these two components: a SMBH is prominent in the LINERs and Sy2s, whereas star formation
increases, as the ${\rm W}2-{\rm W}3$ color becomes redder, along the sequence:
LINER$\rightarrow$Sy2$\rightarrow$TO$\rightarrow$SFG.

Now, when we compare in the MIRDD the BLAGNs with the NELGs, which is impossible to do in the BPT-VO diagram, we
find that the BL Lac objects, the Sy1s and QSOs, fall in the same region of the diagram as the LINERs and Sy2s, their colors forming a sequence that is consistent with the colors predicted by a power law, $F_\nu = \nu^{\alpha}$, where $\alpha$ varies between 0 and $-2$, which is standard for AGNs in the optical-ultraviolet. The MIRDD, therefore, seems like a robust tool to distinguish between AGNs and star-forming galaxies in the nearby universe. In particular, it eliminates the confusion and ambiguities about LINERs, clearly showing that they are genuine AGNs.

Another important characteristic of the MIRDD is the sequence in  ${\rm W}2-{\rm W}3$ colors.  For the NELGs, including the Sy2s, we have clearly shown that this sequence is due to a systematic variation of SFHs with the activity type of the galaxies. Comparing the BLAGNs and BL Lac with the NELGs, we have then found that the color sequence can be extended to: BL~Lac$\rightarrow$LINER$\rightarrow$QSO/Sy1$\rightarrow$Sy2$\rightarrow$TO$\rightarrow$SFG. Our analysis of the SEDs suggests, therefore, that although QSOs, Sy1s and BL Lac have a dominant AGN component, more comparable to LINERs than Sy2s, their differences in ${\rm W}2-{\rm W}3$ are consistent with different levels of star formation in their host galaxies, which would thus decrease along the sequence: BL~Lac$\rightarrow$QSO$\rightarrow$Sy1. This result suggests that the ${\rm W}2-{\rm W}3$ color trace the level of formation in all the emission-line galaxies, irrespective of the importance of the AGN component. 

What is the alternative interpretation for the ${\rm W}2-{\rm W}3$ color sequence? There is a clear trend in the literature to favor an interpretation which assumes that in BLAGNs and BL Lac objects the MIR emission is dominated by the AGN.  This assumption is obviously based on what we see in the optical-ultraviolet part of the spectra. However, according to this assumption, one would naturally expect the MIR colors of these galaxies to follow the AGN luminosity, and the sequence should have been then BL~Lac/QSO$\rightarrow$Sy1$\rightarrow$Sy2$\rightarrow$LINER. The problem here is the LINERs. We have shown that, contrary to what was presented in the literature \citep[c.f.,][]{Wright10}, these AGNs do not experience intense star formation activity, as the ULIRGs, and consequently their typical MIR colors, and SEDs are similar to those of BL~Lac and QSOs, not SFGs, as suggested by the above sequence. The fact that we have two types of LINERs, showing similar optical spectra, but different MIR colors due to their different levels of star formation, is a clear indication that one cannot apply the assumption of a dominant AGN feature in the MIR solely based on their optical-ultraviolet spectra. 

The same remark applies to the BL Lac objects. Based on their optical spectra, BL~Lac are considered to be ``pure'' AGN \citep[e.g.,][]{Plotkin12,Massaro13}. However, we also know that these AGNs are located in non active elliptical galaxies \citep{Kotilainen98,Scarpa00,Falomo14}, which have typical blue colors in the MIR. This is exactly what the MIRDD and SED analysis suggests: they have slightly bluer ${\rm W}2-{\rm W}3$ colors  than the LINERs in the MIRDD, and their SEDs show that it cannot include any star formation components, which is consistent with AGNs in elliptical galaxies. 

Similarly, applying no a priori assumption for the BLAGNs in the MIR, the MIRDD and SED analysis suggest that QSOs and Sy1s are similar kinds of AGNs, where different levels of star formation in their host galaxies explain the variations in the MIR.  Consequently, we believe that the most probable interpretation for the ${\rm W}2-{\rm W}3$ color sequence traced by
emission-line galaxies in the MIRDD is that star formation increases along the sequence
BL~Lac$\rightarrow$LINER$\rightarrow$QSO/Sy1$\rightarrow$Sy2$\rightarrow$TO$\rightarrow$SFG. This interpretation is
consistent with previous studies
\citep[e.g.][]{HamannFerland92,McLeod94,Brotherton99,Page01,Heckman04,Schweitzer06,Wild07,Parra10,Serjeant10,Rafferty11,TP13,Rosario13,Floyd13,Feltre13,Young14,Matsuoka14},
and is in good agreement with theoretical models for the formation of SMBH at the center of galaxies
\citep[e.g.][]{Norman88,Page01,Mouri02,Heckman04,Gurkan04,Cen12}.

We conclude that he possibility to compare BLAGNs with NELGs is an important advantage of the MIRDD. Such a comparison supports the view
that there is a tight connection between the activity type of galaxies in the nearby universe and how these galaxies
form their stars
\citep{Dokuchaev91,HamannFerland92,BoyleTerlevich98,Franceschini99,Haiman04,Menanteau05,LouJiang08,Haiman07,Li07,Bennert08,Fanidakis11,Treister11,Donoso12,TP12a,TP12b,Floyd13,TP13}.

\section{Summary and conclusions} 
\label{CONCLUSION}

In \citet{Coz14}, a new diagnostic diagram in the MIR was proposed, the MIRDD, which is based on WISE data. In the
present study we have performed a stellar population synthesis analysis on the NELGs and compared their MIR colors with those produced by different SED templates. We have also compared the MIRDD of BLAGNs and BL Lac objects with the MIRDD for NELGs, and compared their SEDs. The main results of our study are the following:

\begin{itemize}

\item The MIRDD allows to separate emission-line galaxies according to their activity types, 
by identifying the dominant component in the MIR SED of their host galaxies:  an AGN is predominant in the LINERs, the BL Lac objects, the Sy1s, the Sy2s, and in the QSOs, whereas star formation is more important in the TOs and SFGs. 

\item The LINERs and BL Lac occupy the same region of the MIRDD as the LINERs and Sy2s, consistent with a power law with spectral index varying between 0 and $-2.0$, in good agreement with photoionization model in the optical-ultraviolet. 

\item The ${\rm W}2-{\rm W}3$ color systematically changes in AGNs, in a way which is independent of their luminosities. The most probable interpretation for this change in color is that star formation in the host galaxies of nearby emission-line galaxies increases along the sequence BL~Lac$\rightarrow$LINER$\rightarrow$QSO/Sy1$\rightarrow$Sy2$\rightarrow$TO$\rightarrow$SFG. 

\end{itemize}

Our results underline the importance of star formation in nearby AGNs
\citep[][]{Norman88,HamannFerland92,McLeod94,Brotherton99,Page01,Mouri02,Heckman04,Gurkan04,Schweitzer06,Wild07,
Parra10,Rafferty11,Cen12,TP13,Rosario13,Floyd13,Feltre13,Young14,Matsuoka14}, and 
support the idea that there must
be a tight connection between the formation of their SMBHs and the formation of their host galaxies
\citep{Dokuchaev91,HamannFerland92,BoyleTerlevich98,Franceschini99,Haiman04,Menanteau05,LouJiang08,Haiman07,
Li07,Bennert08,Fanidakis11,Treister11,TP12a,TP12b,Floyd13,TP13}. 

\acknowledgments

The authors acknowledge PROMEP for support through grant 103.5-10-4684, and DAIP for support through grant DAIP-Ugto
(0432/14). They also want to thank an anonymous referee for the many comments and suggestions to improve their study and presentation. This publication makes use of data products from the Wide-field Infrared Survey Explorer, which is a joint project of the University of California, Los Angeles, and the Jet Propulsion Laboratory/California Institute of Technology, funded by the National Aeronautics and Space Administration. The acknowledgement for the SDSS goes to the Alfred P. Sloan Foundation, the Participating Institutions (the full acknowledgement can be found
here: http://www.sdss3.org), the
National Science Foundation, and the U.S. Department of Energy Office of Science. This research has also made use of the NASA/ IPAC Infrared Science Archive, which is operated by the Jet Propulsion Laboratory, California Institute of Technology, under contract with the National Aeronautics and Space Administration, and of the VizieR catalogue access tool, CDS, Strasbourg, France.


\clearpage

\clearpage

\clearpage

\clearpage

\clearpage

\clearpage

\clearpage

\clearpage

\clearpage


\clearpage

\clearpage

\clearpage

\clearpage

\clearpage

\clearpage

\clearpage

\clearpage

\clearpage

\clearpage

\clearpage

\clearpage

\clearpage

\clearpage

\clearpage

\clearpage

\clearpage

\clearpage

\clearpage


\clearpage

\clearpage
\begin{deluxetable}{lccc}
\tabletypesize{\scriptsize}
\tablecaption{Results for the SFH of NELGs \label{tbl-1}}
\tablewidth{0pt}
\tablehead{
\colhead{Activity} & \colhead{Log(SFR$_{\rm Young}$)} & \colhead{Log(SFR$_{\rm Old}$)} & \colhead{Log(t$_{\rm SFMAX}$)} \\
\colhead{Type} & \colhead{(M$_\odot$ yr$^{-1}$)} & \colhead{(M$_\odot$ yr$^{-1}$)} & \colhead{(yrs)}
}
\startdata

LINER   &   -0.16   &  0.66  &   8.71\\
 Sy2      &   0.19   &  0.64   &  9.07\\
 TO2    &     0.33    & 0.55  &   9.20\\
 TO1     &    0.38  &   0.44  &   9.34\\
 SFG2     &   0.50   &  0.28   &  9.37\\
 SFG1    &    0.43   &  0.23   &  9.59\\
\enddata
\end{deluxetable}


\begin{thebibliography}{}

\bibitem[Abazajian et al.(2009)]{Abazajian09} Abazajian, K.~N., Adelman-McCarthy, J.~K., Ag{\"u}eros, M.~A., et al.\ 2009, \apjs, 182, 543 
\bibitem[Assef et al.(2013)]{Assef13} Assef, R.~J., Stern, D., Kochanek, C.~S., et al.\ 2013, \apj, 772, 26 
\bibitem[Baldwin, Phillips \& Terlevich(1981)]{BPT81} Baldwin, J. A., Phillips, M. M. \& Terlevich, R. 1981, \pasp, 93, 5 
\bibitem[Baskin \& Laor(2005)]{Baskin05} Baskin, A. \& Laor, A. 2005, \mnras, 358, 1043 
\bibitem[Barth et al.(1998)]{Barth98} Barth, A.~J., Ho, L.~C., Filippenko, A.~V., \& Sargent, W.~L.~W.\ 1998, \apj, 496, 133
\bibitem[Bertoldi et al.(2003)]{Bertoldi03} Bertoldi, F., Cox, P., Neri, R., et al.\ 2003, \aap, 409, L47
\bibitem[Bennert et al.(2006)]{Bennert06} Bennert, N., Jungwiert, B., Komossa, S., Haas, M. \& Chini, R. 2006, \aap, 446, 919 
\bibitem[Bennert et al.(2008)]{Bennert08} Bennert, N., Canalizo, G., Jungwiert, B., et al.\ 2008, \apj, 677, 846 
\bibitem[Boyle \& Terlevich(1998)]{BoyleTerlevich98} Boyle, B.~J., \& Terlevich, R.~J.\ 1998, \mnras, 293, L49
\bibitem[Brinchmann et al.(2004)]{Brinchmann04} Brinchmann, J., Charlot, S., White, S.~D.~M., et al.\ 2004, \mnras, 351, 1151 
\bibitem[Brotherton et al.(1999)]{Brotherton99} Brotherton, M.~S., van Breugel, W., Stanford, S.~A., et al.\ 1999, \apjl, 520, L87 
\bibitem[Calzetti et al.(2007)]{Calzetti07} Calzetti, D.,  Kennicutt, R.~C., Engelbracht, C.~W., et al.\ 2007, \apj, 666, 870 
\bibitem[Carpineti et al.(2012)]{Carpineti12} Carpineti, A., Kaviraj, S., Darg, D., et al.\ 2012, \mnras, 420, 2139 
\bibitem[Carter et al.(2001)]{Carter01} Carter, B.~J., Fabricant, D.~G., Geller, M.~J., Kurtz, M.~J., \& McLean, B.\ 2001, \apj, 559, 606 
\bibitem[Cen(2012)]{Cen12} Cen, R. 2012, \apj, 755, 28 
\bibitem[Chary \& Elbaz(2001)]{CharyElbaz01} Chary, R., \& Elbaz, D.\ 2001, \apj, 556, 562 
\bibitem[Chen \& Zhang(2006)]{CZ06} Chen, P.~S., \& Zhang, P.\ 2006, \aj, 131, 1942 
\bibitem[Chen et al.(2009)]{Chen09} Chen, X.~Y., Liang, Y.~C., Hammer, F., Zhao, Y.~H., \& Zhong, G.~H.\ 2009, \aap, 495, 457 
\bibitem[Chen et al.(2010)]{Chen10} Chen, X.~Y., Liang, Y.~C., Hammer, F., et al.\ 2010, \aap, 515, A101
\bibitem[Cid Fernandes et al.(2005)]{Cid05} Cid Fernandes, R., Mateus, A., Sodr\'e, L., Stasi\'nska, G. \& Gomes, J.M., 2005, \mnras, 358, 363 
\bibitem[Cid Fernandes et al.(2011)]{CF11} Cid Fernandes, R., Stasi{\'n}ska, G., Mateus, A., \& Vale Asari, N.\ 2011, \mnras, 413, 1687 
\bibitem[Clemens et al.(2013)]{Clemens13} Clemens, M.~S., Negrello, M., De Zotti, G., et al.\ 2013, \mnras, 433, 695
\bibitem[Constantin et al.(2008)]{Constantin08} Constantin, A., Hoyle, F., \& Vogeley, M.~S.\ 2008, \apj, 673, 715 
\bibitem[Constantin et al.(2009)]{Constantin09} Constantin, A., Green, P., Aldcroft, T., et al.\ 2009, \apj, 705, 1336 
\bibitem[Cooke et al.(2000)]{Cooke00} Cooke, A. J., Baldwin, J. A., Ferland, G. J., Netzer, H. \& Wilson, A. S. 2000, \apjs, 129, 517 
\bibitem[Coziol(1996)]{Coz96} Coziol, R.\ 1996, \aap, 309, 345
\bibitem[Coziol et al.(1999)]{Coz99} Coziol, R., Reyes, R.~E.~C., Consid{\`e}re, S., Davoust, E., \& Contini, T.\ 1999,  \aap, 345, 733 
\bibitem[Coziol et al.(2001)]{Coz01} Coziol, R., Doyon, R., \& Demers, S.\ 2001, \mnras, 325, 1081 
\bibitem[Coziol et al.(2011)]{Coz11} Coziol, R., Torres-Papaqui, J.~P., Plauchu-Frayn, I., et al.\ 2011, \rmxaa, 47, 361 
\bibitem[Coziol et al.(2014)]{Coz14} Coziol, R., Torres-Papaqui, J. P., Plauchu-Frayn, I., et al.\ 2014, \rmxaa, 50, 255 
\bibitem[Cutri et al.(2013)]{Cutri13} Cutri, R. M. et al. 2013, VizieR Online Data Catalog: AllWISE Data Release (2013yCat.2328....0C) 
\bibitem[Derry et al.(2003)]{Derry03} Derry, P. M., O’Brien, P. T., Reeves, J. N., et al. 2003, MNRAS, 342, L53
\bibitem[Disney \& Cromwell(1971)]{DC71} Disney, M.~J., \& Cromwell, R.~H.\ 1971, \apjl, 164, L35 
\bibitem[Dokuchaev(1991)]{Dokuchaev91} Dokuchaev, V.~I.\ 1991, \mnras, 251, 564
\bibitem[Donoso et al.(2012)]{Donoso12} Donoso, E., Yan, L., Tsai, C., et al.\ 2012, \apj, 748, 80 
\bibitem[Dopita \& Evans(1986)]{DE86} Dopita, M.~A., \& Evans, I.~N.\ 1986, \apj, 307, 431 
\bibitem[Dopita(1995)]{Dopita95} Dopita, M.~A.\ 1995, \apss, 233, 215 
\bibitem[Dultzin-Hacyan et al.(1988)]{Dultzin88} Dultzin-Hacyan, D., Moles, M. \& Masegosa, J. 1988, \aap, 206, 95 
\bibitem[Falomo et al.(2014)]{Falomo14} Falomo, R., Pian, E.,  \& Treves, A.\ 2014, A\&AR, 22, 73
\bibitem[Fanidakis et al.(2011)]{Fanidakis11} Fanidakis, N., Baugh, C.~M., Benson, A.~J., et al.\ 2011, \mnras, 410, 53 
\bibitem[Feltre et al.(2013)]{Feltre13} Feltre, A., Hatziminaoglou, E., Hern{\'a}n-Caballero, A., et al.\ 2013, \mnras, 434, 2426 
\bibitem[Ferland et al.(1998)]{Ferland98} Ferland, G.~J., Korista, K.~T., Verner, D.~A., et al.\ 1998, \pasp, 110, 761 
\bibitem[Flohic et al.(2006)]{Flohic06} Flohic, H.~M.~L.~G., Eracleous, M., Chartas, G., Shields, J.~C., \& Moran, E.~C.\ 2006, \apj, 647, 140 
\bibitem[Floyd et al.(2013)]{Floyd13} Floyd, D.~J.~E., Dunlop, J.~S., Kukula, M.~J., et al.\ 2013, \mnras, 429, 2 
\bibitem[Forbes(1993)]{Forbes93} Forbes, D.~A.\ 1993, \apss, 205, 37 
\bibitem[Franceschini et al.(1999)]{Franceschini99} Franceschini, A., Hasinger, G., Miyaji, T., \& Malquori, D.\ 1999, \mnras, 310, L5 
\bibitem[Giraud et al.(2011)]{Giraud11} Giraud, E., Gu, Q.-S., Melnick, J., et al.\ 2011, Research in Astronomy and Astrophysics, 11, 245 
\bibitem[Gon\c{c}alves et al.(1999)]{GVV99} Gon\c{c}alves, A. C., V\'eron-Cetty \& M.-P., V\'eron, P. 1999, \aaps, 135, 437 
\bibitem[Gonz{\'a}lez Delgado et al.(2008)]{GD08} Gonz{\'a}lez Delgado, R.~M., P{\'e}rez, E., Cid Fernandes, R., \& Schmitt, H.\ 2008, \aj, 135, 747 
\bibitem[Goodrich \& Keel(1986)]{GK86} Goodrich, R.~W., \& Keel, W.~C.\ 1986, \apj, 305, 148 
\bibitem[G\"urkan, Freitag \& Rasio(2004)]{Gurkan04} G\"urkan, M. A., Freitag, M. \& Rasio, F. A. 2004, \apj, 604, 632 
\bibitem[Haan et al.(2008)]{Haan08} Haan, S., Schinnerer, E., Mundell, C.~G., Garc{\'{\i}}a-Burillo, S., \& Combes, F.\ 2008, \aj, 135, 232 
\bibitem[Haiman et al.(2004)]{Haiman04} Haiman, Z., Ciotti, L., \& Ostriker, J.~P.\ 2004, \apj, 606, 763
\bibitem[Haiman et al.(2007)]{Haiman07} Haiman, Z., Jimenez, R., \& Bernardi, M.\ 2007, \apj, 658, 721 
\bibitem[Hamann \& Ferland(1992)]{HamannFerland92} Hamann, F., \& Ferland, G.\ 1992, \apjl, 391, L53 
\bibitem[Hamann \& Ferland(1993)]{Hamann93} Hamann, F. \& Ferland, G. 1993, \apj, 418, 11 
\bibitem[Hanami et al.(2012)]{Hanami12} Hanami, H., Ishigaki,  T., Fujishiro, N., et al.\ 2012, \pasj, 64, 70 
\bibitem[Heckman(1980)]{Heckman80} Heckman, T.~M.\ 1980, \aap, 87, 152 
\bibitem[Heckman(1986)]{Heckman86} Heckman, T.~M.\ 1986, \pasp, 98, 159 
\bibitem[Heckman et al.(2004)]{Heckman04} Heckman, T. M., Kauffmann, G., Brinchmann, J., Charlot, S., Tremonti, C. \& White, S. D. M. 2004, \apj, 613, 109 
\bibitem[Herberich, Sikorski \& Hothorn(2010)]{HSH10} Herberich, E., Sikorski, J., Hothorn, T. 2010, PLoS ONE 5(3): e9788.doi:10.1371/journal.pone.0009788 
\bibitem[Ho et al.(1993)]{HFS93} Ho, L.~C., Filippenko, A.~V., \& Sargent, W.~L.~W.\ 1993, \apj, 417, 63
\bibitem[Ho(1996)]{Ho96} Ho, L.~C.\ 1996, \pasp, 108, 637 
\bibitem[Ho et al.(1997)]{HFS97} Ho, L.~C., Filippenko, A.~V., \& Sargent, W.~L.~W.\ 1997, \apj, 487, 568 
\bibitem[Ho(1999)]{Ho99} Ho, L.~C.\ 1999, Advances in Space Research, 23, 813
\bibitem[Ho, Filippenko \& Sargent(2003)]{HFS03} Ho, L. C., Filippenko, A. V., Sargent, W. L. W. 2003, \apj, 583, 159
\bibitem[Hothorn, Bretz \& Westfall(2008)]{HBW08} Hothorn, T., Bretz, F., Westfall, P., 2008, Biom J, 50, 346
\bibitem[Jarrett et al.(2011)]{Jarrett11} Jarrett, T.~H., Cohen, M., Masci, F., et al.\ 2011, \apj, 735, 112
\bibitem[Jarrett et al.(2013)]{Jarrett13} Jarrett, T.~H., Masci, F., Tsai, C.~W., et al.\ 2013, \aj, 145, 6
\bibitem[Kauffmann et al.(2003)]{Kau03} Kauffmann, G. et al., 2003, \mnras, 346, 105 
\bibitem[Kauffmann(2009)]{Kau09} Kauffmann, G. 2009, \aap, 500, 201 
\bibitem[Kennicutt(1992)]{Kennicutt92} Kennicutt, R. C. Jr., 1992, \apjs, 79, 255
\bibitem[Kennicutt et al.(2003)]{Kennicutt03} Kennicutt, R.~C.,  Jr., Armus, L., Bendo, G., et al.\ 2003, \pasp, 115, 928 
\bibitem[Kewley et al.(2001)]{Kew01} Kewley, L.J., Dopita, M.A., Sutherland, R.S., Heisler, C.A. Trevena, J., 2001, \apj, 556, 121 
\bibitem[Kewley et al.(2006)]{Kew06} Kewley, L.J., Groves, B., Kauffmann, G., Heckman, T., 2006, \mnras, 372, 961 
\bibitem[Kim et al.(1998)]{Kim98} Kim, D.-C., Veilleux, S., \& Sanders, D.~B.\ 1998, \apj, 508, 627
\bibitem[Kotilainen et  al.(1998)]{Kotilainen98} Kotilainen, J.~K., Falomo, R., \& Scarpa, R.\ 1998, \aap, 336, 479 
\bibitem[Larkin et al.(1998)]{Larkin98} Larkin, J.~E., Armus, L., Knop, R.~A., Soifer, B.~T., \& Matthews, K.\ 1998, \apjs, 114, 59 
\bibitem[Laurent et  al.(2000)]{Laurent00} Laurent, O., Mirabel, I.~F., Charmandaris, V., et al.\ 2000, \aap, 359, 887 
\bibitem[Lawrence et al.(1985)]{Lawrence85} Lawrence, A., Ward, M., Elvis, M., et al. 1985, \apj, 291, 117 
\bibitem[Lee et al.(2007)]{Lee07} Lee, J.~H., Lee, M.~G., Kim, T., et al.\ 2007, \apjl, 663, L69 \
\bibitem[Leipski et al.(2014)]{Leipski14} Leipski, C., Meisenheimer, K., Walter, F., et al.\ 2014, \apj, 785, 154 
\bibitem[Li et al.(2007)]{Li07} Li, Y., Hernquist, L., Robertson, B., et al.\ 2007, \apj, 665, 187 
\bibitem[Lou \& Jiang(2008)]{LouJiang08} Lou, Y.-Q., \& Jiang, Y.-F.\ 2008, \mnras, 391, L44
\bibitem[Lynden-Bell(1969)]{Lynden-Bell69} Lynden-Bell, D.\ 1969, \nat, 223, 690 
\bibitem[McLeod \& Rieke(1994)]{McLeod94} McLeod, K. K. \& Rieke, G. H. 1994, \apj, 431, 137 
\bibitem[Maoz et al.(1998)]{Maoz98} Maoz, D., Koratkar, A., Shields, J.~C., et al.\ 1998, \aj, 116, 55 
\bibitem[Maoz(1999)]{Maoz99} Maoz, D.\ 1999, Advances in Space Research, 23, 855 
\bibitem[Maoz(2007)]{Maoz07} Maoz, D.\ 2007, \mnras, 377, 1696 
\bibitem[Maoz(2008)]{Maoz08} Maoz, D.\ 2008, Journal of Physics Conference Series, 131, 012036
\bibitem[Masegosa et al.(2011)]{Masegosa11} Masegosa, J., M{\'a}rquez, I., Ramirez, A., \& Gonz{\'a}lez-Mart{\'{\i}}n, O.\ 2011, \aap, 527, A23 
\bibitem[Massaro et al.(2011)]{Massaro11} Massaro, F., D'Abrusco, R., Ajello, M., Grindlay, J.~E., \& Smith, H.~A.\ 2011, \apjl, 740, L48
\bibitem[Massaro et al.(2013)]{Massaro13} Massaro, F., Paggi, A., Errando, M., et al.\ 2013, \apjs, 207, 16 
\bibitem[Mateos et al.(2012)]{Mateos12} Mateos, S., Alonso-Herrero, A., Carrera, F.~J., et al.\ 2012, \mnras, 426, 3271
\bibitem[Matsuoka et al.(2014)]{Matsuoka14} Matsuoka, Y., Strauss, M.~A., Price, T.~N., III, \& DiDonato, M.~S.\ 2014, \apj, 780, 162 
\bibitem[McCall et al.(1985)]{McCall85} McCall, M.~L., Rybski, P.~M., \& Shields, G.~A.\ 1985, \apjs, 57, 1 
\bibitem[Menanteau et al.(2005)]{Menanteau05} Menanteau, F., Martel, A.~R., Tozzi, P., et al.\ 2005, \apj, 620, 697
\bibitem[Montero-Dorta et al.(2009)]{Montero-Dorta09} Montero-Dorta, A.~D., Croton, D.~J., Yan, R., et al.\ 2009, \mnras, 392, 125 
\bibitem[Mouri \& Taniguchi(2002)]{Mouri02} Mouri, H. \& Taniguchi, Y. 2002, \apj, 566, L17
\bibitem[Murakami et al.(2007)]{Murakami07} Murakami, H., Baba, H., Barthel, P., et al.\ 2007, \pasj, 59, 369
\bibitem[Nagao, Maiolino \& Marconi(2006)]{Nagao06} Nagao, T., Maiolino, R. \& Marconi, A. 2006, \aap, 447, 863
\bibitem[Netzer(2009)]{Netzer09} Netzer, H.\ 2009, \mnras, 399, 1907 
\bibitem[Norman \& Scoville(1988)]{Norman88} Norman, C. \& Scoville, N. 1988, \apj, 332, 124 
\bibitem[Ochsenbein, Bauer \& Marcout(2000)]{OBM00} Ochsenbein, F., Bauer, P. \& Marcout, J. 2000, \aaps, 143, 23 
\bibitem[Osterbrock(1970)]{Osterbrock70} Osterbrock, D. E., 1970, QJRAS, 11, 1990
\bibitem[Osterbrock \& Dahari(1983)]{OD83} Osterbrock, D.~E., \& Dahari, O.\ 1983, \apj, 273, 478
\bibitem[Osterbrock \& Shaw(1988)]{OS88} Osterbrock, D.~E., \& Shaw, R.~A.\ 1988, \apj, 327, 89 
\bibitem[Osterbrock \& Ferland(2006)]{OF06} Osterbrock, D.~E., \& Ferland, G.~J.\ 2006, Astrophysics of gaseous nebulae and active galactic nuclei, 2nd.~ed., University Science Books
\bibitem[Page et al.(2001)]{Page01} Page, M. J., Stevens, J. A., Mittaz, J. P. D. \& Carrera, F. J. 2001, Sci, 294, 2516
\bibitem[Papaderos et al.(2013)]{Papaderos13} Papaderos, P., Gomes, J.~M., V{\'{\i}}lchez, J.~M., et al.\ 2013, \aap, 555, L1 
\bibitem[Parra et al.(2010)]{Parra10} Parra, R., Conway, J. E., Aalto, S., Appleton, P. N., Norris, R. P., Pihlstr\"om, Y. M. \& Kewley, L. J. 2010, \apj, 720, 555 
\bibitem[Plauchu-Frayn et al.(2012)]{Plauchu12} Plauchu-Frayn, I., Del Olmo, A., Coziol, R., \& Torres-Papaqui, J.~P.\ 2012, \aap, 546, A48 
\bibitem[Plotkin et al.(2012)]{Plotkin12} Plotkin, R.~M.,  Anderson, S.~F., Brandt, W.~N., et al.\ 2012, \apjl, 745, LL27
\bibitem[Polletta et al.(2007)]{Polletta07} Polletta, M., Tajer, M., Maraschi, L., et al.\ 2007, \apj, 663, 81 
\bibitem[Rafferty et al.(2011)]{Rafferty11} Rafferty, D.~A., Brandt, W.~N., Alexander, D.~M., et al.\ 2011, \apj, 742, 3 
\bibitem[Rawlings \& Saunders(1991)]{RawlingsSaunders91} Rawlings, S., \& Saunders, R.\ 1991, \nat, 349, 138 
\bibitem[Rosario et al.(2013)]{Rosario13} Rosario, D.~J., Burtscher, L., Davies, R., et al.\ 2013, \apj, 778, 94 
\bibitem[Roseboom et al.(2013)]{Roseboom13} Roseboom, I.~G., Lawrence, A., Elvis, M., et al.\ 2013, \mnras, 429, 1494 
\bibitem[Sarzi et al.(2005)]{Sarzi05} Sarzi, M.,Rix, H.-W., Shields, J.~C., et al.\ 2005, \apj, 628, 169 
\bibitem[Serjeant \& Hatziminaoglou(2009)]{Serjeant09} Serjeant, S., \& Hatziminaoglou, E.\ 2009, \mnras, 397, 265 
\bibitem[Serjeant et al.(2010)]{Serjeant10} Serjeant, S., Bertoldi, F., Blain, A.~W., et al.\ 2010, \aap, 518, L7 
\bibitem[Schmitt et al.(1997)]{Schmitt97} Schmitt, H.~R., Kinney, A.~L., Calzetti, D., \& Storchi Bergmann, T.\ 1997, \aj, 114, 592
\bibitem[Sheth et al.(2010)]{Sheth10} Sheth, K., Regan, M., Hinz, J.~L., et al.\ 2010, \pasp, 122, 1397 
\bibitem[Shields(1992)]{Shields92} Shields, J.~C.\ 1992, \apjl, 399, L27 
\bibitem[Singh et al.(2013)]{Singh13} Singh, R., van de Ven, G., Jahnke, K., et al.\ 2013, \aap, 558, A43 
\bibitem[Schweitzer et al.(2006)]{Schweitzer06} Schweitzer, M. et al. 2006, \apj, 649, 79
\bibitem[Soltan(1982)]{Soltan82} Soltan, A. 1982, MNRAS, 200, 115 
\bibitem[Scarpa et al.(2000)]{Scarpa00} Scarpa, R., Urry, C.~M.,  Padovani, P., Calzetti, D., \& O'Dowd, M.\ 2000, \apj, 544, 258 
\bibitem[Stasi{\'n}ska et al.(2006)]{Stasinska06} Stasi{\'n}ska, G., Cid Fernandes, R., Mateus, A., Sodr{\'e}, L., \& Asari, N.~V.\ 2006, \mnras, 371, 972
\bibitem[Stasi{\'n}ska et al.(2008)]{S08} Stasi{\'n}ska, G., Vale Asari, N., Cid Fernandes, R., et al.\ 2008, \mnras, 391, L29 
\bibitem[Starling et al.(2005)]{Starling05} Starling, R.~L.~C., Page, M.~J., Branduardi-Raymont, G., et al.\ 2005, \mnras, 356, 727 
\bibitem[Stern et al.(2012)]{Stern12} Stern, D., Assef, R.~J., Benford, D.~J., et al.\ 2012, \apj, 753, 30 
\bibitem[Storchi-Bergmann \& Pastoriza(1989)]{Storchi89} Storchi Bergmann, T. \& Pastoriza, M. G., 1989, \apj, 347, 195 
\bibitem[Storchi-Bergmann(1991)]{Storchi91} Storchi Bergmann, T. 1991, MNRAS, 249, 404 
\bibitem[Stoughton et al.(2002)]{Stoughton02} Stoughton, C., Lupton, R. H., Bernardi, M. et al., 2002, \aj, 123, 485 
\bibitem[Taniguchi et al.(2000)]{Taniguchi00} Taniguchi Y., Shioya Y. \& Murayama T. 2000, \aj, 120, 1265 
\bibitem[Terashima et al.(2000)]{Terashima00} Terashima, Y., Ho, L.~C., Ptak, A.~F., et al.\ 2000, \apj, 533, 729 
\bibitem[Terlevich \& Melnick(1985)]{TM85} Terlevich, R., \& Melnick, J.\ 1985, \mnras, 213, 841 
\bibitem[Torres-Papaqui et al.(2011)]{TP11} Torres-Papaqui, J.~P., Coziol, R., \& Ortega-Minakata, R.~A.\ 2011, Acta Universitaria, 21, 82 
\bibitem[Torres-Papaqui et al.(2012a)]{TP12a} Torres-Papaqui, J.~P., Coziol, R., Andernach, H., et al.\ 2012a, \rmxaa, 48, 275 
\bibitem[Torres-Papaqui et al.(2012b)]{TP12b} Torres-Papaqui, J.~P., Coziol, R., Ortega-Minakata, R.~A., \& Neri-Larios, D.~M.\ 2012b, \apj, 754, 144 
\bibitem[Torres-Papaqui et al.(2013)]{TP13} Torres-Papaqui, J.~P., Coziol, R., Plauchu-Frayn, I., Andernach, H., Ortega-Minakata, R.~A., 2013, \rmxaa, 49, 311 
\bibitem[Treister et al.(2011)]{Treister11} Treister, E., Schawinski, K., Volonteri, M., Natarajan, P., \& Gawiser, E.\ 2011, \nat, 474, 356 
\bibitem[Veilleux \& Osterbrock(1987)]{VO87} Veilleux, S. \& Osterbrock, D.E., 1987, \apjs, 63, 295 
\bibitem[Veilleux et al.(1995)]{Veilleux95} Veilleux, S., Kim, D.-C., Sanders, D.~B., Mazzarella, J.~M., \& Soifer, B.~T.\ 1995, \apjs, 98, 171 
\bibitem[Veilleux et al.(1999)]{Veilleux99} Veilleux, S., Kim, D.-C., \& Sanders, D.~B.\ 1999, \apj, 522, 113 
\bibitem[V{\'e}ron-Cetty \& V{\'e}ron(2010)]{VCV10} V{\'e}ron-Cetty, M.-P., \& V{\'e}ron, P.\ 2010, \aap, 518, A10 
\bibitem[Viegas \& de Gouveia dal Pino(1992)]{Viegas92} Viegas, S. M. \& de Gouveia dal Pino, E. M. 1992, ApJ, 384, 467 
\bibitem[V\'eron, Gon\c{c}alves \& V\'eron-Cetty(1997)]{VGV97} V\'eron, P., Gon\c{c}alves, A. C. \& V\'eron-Cetty, M.-P. 1997, \aa, 319, 52
\bibitem[Vila-Vilaro(2000)]{Vila-Vilaro00} Vila-Vilaro, B.\ 2000, \pasj, 52, 305 
\bibitem[Wang \& Wei(2008)]{WW08} Wang, J., \& Wei, J.~Y.\ 2008, \apj, 679, 86 
\bibitem[Wang et al.(2008)]{Wang08} Wang, R., Carilli, C.~L.,  Wagg, J., et al.\ 2008, \apj, 687, 848
\bibitem[Wang et al.(2010)]{Wang10} Wang, R., Carilli, C.~L.,  Neri, R., et al.\ 2010, \apj, 714, 699 
\bibitem[Wild et al.(2007)]{Wild07} Wild, V., Kauffmann, G., Heckman, T., et al.\ 2007, \mnras, 381, 543 
\bibitem[Wild, Heckman \& Charlot(2010)]{Wild10} Wild, V., Heckman, T. \& Charlot, S. 2010, \mnras, 405, 933 
\bibitem[Willner et al.(1985)]{Willner85} Willner, S.~P., Elvis, M., Fabbiano, G., Lawrence, A., \& Ward, M.~J.\ 1985, \apj, 299, 443 
\bibitem[Wright et al.(2010)]{Wright10} Wright, E. L. et al., 2010, AJ, 140, 1868 
\bibitem[Wu et al.(1998)]{Wu98} Wu, H., Zou, Z.~L., Xia, X.~Y., \& Deng, Z.~G.\ 1998, \aaps, 132, 181 
\bibitem[Xu et al.(1998)]{Xu98} Xu, C., Hacking, P.~B., Fang, F., et al.\ 1998, \apj, 508, 576 
\bibitem[Yan et al.(2013)]{Yan13} Yan, L., Donoso, E., Tsai, C.-W., et al.\ 2013, \aj, 145, 55 
\bibitem[Yan \& Blanton(2012)]{YB12} Yan, R., \& Blanton, M.~R.\ 2012, \apj, 747, 61 
\bibitem[York et al.(2000)]{York00} York, D. G., Adelman, J., Anderson, J.E., Jr. et al., 2000, \aj, 120, 1579
\bibitem[Young et al.(2014)]{Young14} Young, J.~E., Eracleous, M., Shemmer, O., et al.\ 2014, \mnras, 438, 217
\bibitem[Yuan et al.(2010)]{Yuan10} Yuan, T.-T., Kewley, L.~J., \& Sanders, D.~B.\ 2010, \apj, 709, 884



\end{thebibliography}
\end{document}